\documentclass[aps,prb,twocolumn]{revtex4-1}
\pdfoutput=1
\usepackage[latin9]{inputenc}
\usepackage{graphicx,epsf}
\usepackage{amssymb}
\usepackage{amsmath}
\usepackage{color, soul}
\usepackage{bm}
\usepackage[caption=false]{subfig}
\usepackage{placeins}

\newcommand{\w}{\omega}
\newcommand{\eps}{\epsilon}
\newcommand{\HMF}{\mathcal{H}_{\rm MF}}
\newcommand{\TN}{T_{\rm N}}
\newcommand{\Mst}{M_{\rm st}}
\newcommand{\hst}{h_{\rm st}}
\newcommand{\hc}{h_{\rm c}}
\newcommand{\hcz}{h_{\rm c}^0}
\newcommand{\qcz}{q_{\rm c}^0}
\newcommand{\Del}{\Delta_{\rm el}}
\newcommand{\Tel}{T_{\rm el}}
\newcommand{\Dnu}{\Delta_{\rm nu}}
\newcommand{\Tnu}{T_{\rm nu}}
\newcommand{\Wnu}{W_{\rm nu}}
\newcommand{\Eps}{E_{\rm ps}}
\newcommand{\lhf}{LiHoF$_{4}$}

\RequirePackage[
   hyperindex,colorlinks,bookmarksnumbered,
   plainpages=true, pdfstartview=FitH,bookmarks=false
]{hyperref}
\hypersetup{linkcolor=blue,urlcolor=blue,citecolor=blue}

\definecolor{darkgreen}{rgb}{0,0.5,0}
\definecolor{darkblue}{rgb}{0,0,0.5}
\definecolor{purple}{rgb}{0.35,0,0.35}
\definecolor{orange}{rgb}{0.9,0.4,0}

\graphicspath{{./}{./figures/}}

\begin{document}
\title{
Limits to magnetic quantum criticality from nuclear spins
}

\author{Heike Eisenlohr}
\author{Matthias Vojta}
\affiliation{Institut f\"ur Theoretische Physik and W\"urzburg-Dresden Cluster of Excellence ct.qmat, Technische Universit\"at Dresden,
01062 Dresden, Germany}

\date{\today}

\begin{abstract}
The phenomenology of quantum phase transitions concerns physics at low temperatures and energies, and corresponding solid-state experiments often reach millikelvin temperatures. However, this is a scale where in many solids the influence of nuclear spins and their hyperfine interaction is no longer negligible. This may limit the observability of electronic quantum critical phenomena.
Here we discuss how continuous magnetic quantum phase transitions get influenced, modified, or destroyed by the coupling to nuclear spins. We use simple yet paradigmatic spin models for magnetic quantum criticality and determine modifications to the phase diagram, the excitation spectrum, and thermodynamics due to the presence of nuclear spins. We estimate crossover scales below which purely electronic quantum criticality is no longer observable, and discuss the distinct physics emerging at low temperatures. Our results are relevant for a variety of compounds displaying magnetic quantum phase transitions and, more generally, highlight the sensitivity of quantum critical systems to small perturbations.
\end{abstract}

\maketitle


\section{Introduction}

Quantum criticality \cite{ssbook,mv_rop,hvl} in solids continues to be a research field of wide interest for a number of reasons:
Quantum criticality comes with unusual phenomena not seen in stable phases, for instance, quantum criticality is likely to host the key to the understanding of a variety of fascinating strange-metal regimes.\cite{Varma16, Hussey18} Conceptually, quantum critical points have proven to be viable starting points for the understanding of complex phase diagrams. Last but not least, entirely novel phases may nucleate in the vicinity of quantum phase transitions, most prominently unconventional superconductivity. On the theory front, recent developments include an improved understanding of quantum criticality in metals and semimetals,\cite{Schattner16, Brando16, Bera16} detailed investigations of deconfined quantum criticality \cite{You16, Wang16} and related ideas on field-theoretic dualities,\cite{Senthil19} as well as the study of quench dynamics across quantum critical points.\cite{Heyl17, Dziarmaga10, Heyl13}

In condensed-matter experiments, quantum phase transitions have been studied in local-moment insulators, metals, and superconductors. In many of these systems, microscopic energy scales are of order $10$\,K or below, such that reaching the asymptotic critical regime requires temperatures significantly below $1$\,K. Indeed, state-of-the-art experiments probing quantum criticality routinely reach temperatures of $20$\,mK and below. At such low temperatures, a number of perturbing factors become important which are often neglected in the standard theoretical modelling, for instance small symmetry-breaking terms or crystalline defects.
In magnetic systems, nuclear spins play a particular role, as they constitute additional degrees of freedom which can actively couple to the critical modes of the quantum phase transition. This will modify the electronic criticality of the host system, and two questions are pertinent:
(i) What is the temperature or energy scale below which the influence of the nuclear spins can no longer be neglected?
(ii) What happens below this scale, i.e., what is the fate of the quantum critical point at lowest temperatures?
In the existing literature, only a few concrete cases have been looked at. One is the transverse-field Ising magnet \lhf,\cite{bitko96} which combines a relatively small coupling between electronic spins with a sizeable hyperfine coupling and large nuclear moments. Here it has been established that nuclear spins produce a significant shift of the critical field,\cite{bitko96} and their influence on excitation modes has also been investigated.\cite{Ronnow05}
Given that numerous (potentially) magnetic ions host a sizeable nuclear moment, for instance Nb, Ho, In, V, Sc, Co and Pr,\cite{TableNuclMoments} a more general view concerning both questions (i) and (ii) is highly desirable.

It is the purpose of this paper to close this gap. Starting from symmetry considerations, we discuss different scenarios for the fate of magnetic quantum phase transitions if nuclear spins are coupled to electrons.
Field-driven order--disorder transitions generically continue to exist, but get shifted by nuclear hyperfine coupling. In contrast, phase transitions in the absence of an external magnetic field are more subtle, because the decoupled nuclear spins constitute a highly degenerate mani\-fold of states: If an electronic phase preserves time-reversal symmetry, this symmetry may be spontaneously broken upon introducing hyperfine coupling, i.e., weak magnetic order driven by nuclear spins may occur at low temperatures.
As a result, a pressure-driven electronic order--disorder transition can disappear entirely, i.e., get smeared. For both field-driven and pressure-driven cases, we present model calculations illustrating the modifications and distinct crossover scales induced by nuclear spins, and we determine the behavior of thermodynamic observables and of the excitation spectrum near the (putative) quantum critical point.
We also extend the considerations to frustrated systems and briefly discuss the fate of certain transitions involving spin-liquid states and of deconfined quantum critical points under the influence of nuclear hyperfine coupling.
Our predictions for quantum critical materials with nuclear spins can be verified using spectroscopic techniques, such as inelastic neutron scattering, as well as thermodynamic measurements, e.g., of the specific heat.

We note that the appearance of nuclear spin order in an otherwise disordered phase has been discussed first for metals in Ref.~\onlinecite{Froehlich40} and more recently for one-dimensional Luttinger liquids, as realized in carbon nanotubes and other quantum wires,\cite{Braunecker09, Braunecker09b, Scheller14} as well as for two-dimensional electron gases.\cite{Simon07}
The effect of nuclear spins on the excitations of ordered magnets has seen extensive discussions, with early works in Refs.~\onlinecite{Sherrington70, Sherrington73}.


\subsection{Outline}

The remainder of the paper is organized as follows:
Sec.~\ref{sec:symm} starts with general considerations on symmetries and the question how phases may get modified by the presence of nuclear spins.
Sec.~\ref{sec:ising} is devoted to field-driven magnetic transitions, with concrete calculations for the transverse-field Ising model. Pressure-driven magnetic transitions are subject of Sec.~\ref{sec:dimers}, and we discuss results for a particular model of coupled dimers.
Our analysis also shows that additional phases and phase transitions may be introduced by hyperfine coupling, and we illustrate this in Sec.~\ref{sec:novel}.
Finally, the influence of hyperfine coupling on strongly frustrated systems, such as spin liquids, and their quantum phase transitions are briefly discussed in Sec.~\ref{sec:exotic}.
A discussion of broader implications of our results closes the paper. Technical details of the calculations are relegated to the appendices.


\section{Symmetries and stability of phases}
\label{sec:symm}

The goal of this paper is to discuss the fate of quantum phase transitions and their associated quantum critical regime upon including the coupling to nuclear spins, but in order to do so, one first needs to discuss the fate of the stable phases involved. As we argue below, this issue is far from trivial.
Here, we will be exclusively concerned with phase transitions involving, at the microscopic level, magnetic degrees of freedom. The arguments will be illustrated using local-moment models appropriate for magnetic insulators; most considerations will also apply, at least qualitatively, to metals and their magnetic quantum phase transitions.

We will assume the local electronic degrees of freedom to be spins $\vec{S}$. More generally, one needs to consider the manifold of the lowest crystal-field states of the active ions and their magnetic moments. For non-Kramers ions the ground-state manifold cannot be described as a standard angular momentum, which may lead to large quantitative differences compared to a description with effective spins. One case in point is Ho, and we will come back to this in Sec.~\ref{sec:LHF} below.

The hyperfine interaction between electronic moments $\vec S$ and nuclear moments $\vec I$, a sum of dipolar interaction and Fermi contact interaction, is local and in general anisotropic, $H_{\rm hf} = \vec{S} A \vec{I}$ where $A$ is a $3\times3$ matrix, with the symmetries dictated by the electronic states and hence the crystalline symmetry. For simplicity, we will neglect anisotropies of $H_{\rm hf}$, and we will therefore replace it by $H_{\rm hf} = A \vec{S} \cdot \vec{I}$, with antiferromagnetic $A>0$ . In any case, the hyperfine interaction will not have lower symmetry than the electronic spin-spin interaction.

To get started, it is useful to consider the hyperfine interaction as a perturbation to the model of electronic interacting spins. In the absence of an external magnetic field, this perturbative treatment expands about \emph{free} nuclear spins, i.e., a highly degenerate many-body state present at $A=0$. The hyperfine coupling $A$ to nuclear spins can therefore constitute a \emph{singular} perturbation to a given electronic state. As we will illustrate below, it can induce spontaneous magnetic order in an otherwise quantum paramagnetic state. \cite{Froehlich40, Simon07, Braunecker09, Braunecker09b}
This is different in the presence of a magnetic field because then the nuclear spins are polarized at $A=0$ via the nuclear Zeeman coupling, and perturbation theory in $A$ is regular. Finally, an electronic state with dipolar magnetic order present at $A=0$ will imprint its order on the nuclear spins if a small hyperfine coupling is switched on.

For magnetic order--disorder quantum phase transitions which are driven by an external magnetic field, we therefore conclude that the two phases remain qualitatively intact in the presence of nuclear spins. However, hyperfine-induced renormalizations will in general lead to quantitative modifications and in particular to a shift of the quantum phase transition point. Detailed considerations for a concrete model are in Sec.~\ref{sec:ising}.

In contrast, for order--disorder transitions in the absence of an external field, e.g., driven by pressure, it is possible that the disordered phase ceases to exist in the low-temperature limit due to nuclear-spin-induced order. Then, the quantum phase transition either disappears entirely, i.e., is smeared into a crossover as illustrated in Sec.~\ref{sec:dimers}, or it may become a transition between two different ordered states.

This discussion shows that, in general, the presence or absence of time reversal symmetry in the nominally disordered phase determines whether a quantum phase transition is smeared or shifted upon inclusion of nuclear spins. An exception are phases with broken time reversal but no dipolar order, such as chiral spin liquids, which may display spontaneous nuclear spin order, such that a transition into an adjacent magnetic state gets smeared.

More complicated situations occur if both involved electronic phases display symmetry-breaking order or if one or both of phases are topological in nature; the latter also concerns the interesting question of the fate of a quantum spin liquid upon coupling it to nuclear spins. Finally, one can envision rare cases where nuclear spins lead to the emergence of new phases near the original electronic quantum critical point. A qualitative discussion of these ideas is in Sec.~\ref{sec:novel} and \ref{sec:exotic}.


\section{Shifted transitions: Transverse-field Ising magnet}
\label{sec:ising}

\begin{figure}[tb]
	\centering
	\includegraphics[width=0.7\columnwidth]{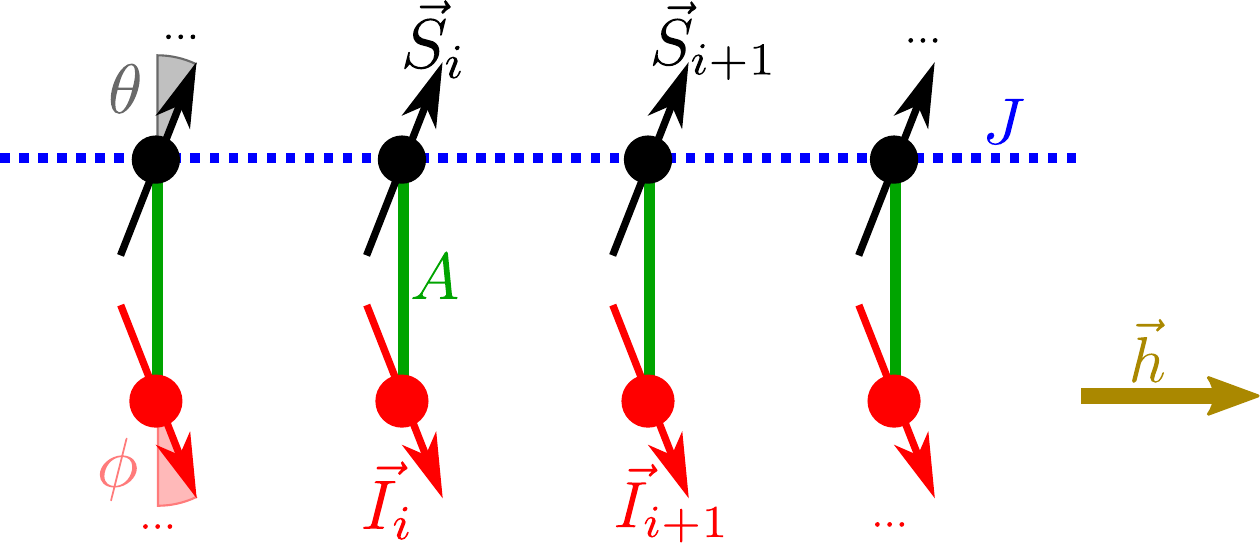}
	\caption{
Transverse-field Ising model with nuclear spins (schematic): electronic spins (black arrows) are coupled by a nearest-neighbor Ising interaction $J$ (blue dashed lines). The hyperfine coupling $A$ (green lines) links them to the nuclear spins (red arrows). The angles $\theta (\phi$) quantify the deviation from the zero-field state for the electronic (nuclear) spins. The model can be generalized to any space dimension.
}
	\label{fig:ising-model}
\end{figure}

As explained above, quantum phase transitions where time-reversal symmetry is broken in both phases can be expected to survive -- and get shifted -- upon including nuclear spins. This applies in particular to transitions driven by an external magnetic field. As an example we consider the well-known transverse-field Ising model augmented by hyperfine coupling to nuclear spins, see Fig.~\ref{fig:ising-model}. It is described by the Hamiltonian
\begin{align}
	\mathcal{H}^{\rm TI} &= - J \sum_{\langle i j \rangle} S_{iz} S_{jz} - \vec H \cdot \sum_i \left( g_e \vec S_i +g_N \vec I_i \right) \notag\\
&+ A \sum_i \vec S_i \cdot \vec I_i
	\label{eq:ising-Ham}
\end{align}
where the summation in the first term runs over pairs $\langle i j \rangle$ of nearest-neighbor sites on a regular lattice, and $\vec S_i$ ($\vec I_i$) are the electronic (nuclear) spins on site $i$, respectively.
The electronic and nuclear $g$ tensors, $g_e$ and $g_N$, are assumed to be isotropic. Both depend on the material under consideration; their ratio can be expected to be of the order of $g_e/g_N \sim 10^3$, resulting from the mass ratio of electron and nucleon. In the following, we will absorb $g_e$ in the definition of the external field, $\vec h = g_e \vec H$ and write the nuclear Zeeman coupling as $\tilde{g}_N \vec{h} \cdot \sum_i \vec{I}_i$ where we take $\tilde{g}_N=10^{-3}$ unless noted otherwise.
Concrete results will be shown for a cubic-lattice model, with spin sizes $S=I=1/2$ and $J=1$ as unit of energy.

\subsection{General considerations}

In the absence of nuclear spins, the electronic system can be driven through a phase transition  from a symmetry-breaking easy-axis ferromagnet at small transverse fields $\vec{h} = h \hat{e}_x$ to a field-polarized paramagnetic state at large $h$. Both phases are gapped, with the gap vanishing at the QPT. At the mean-field level the phase transition occurs at $h_c=2dSJ$, with $2d=6$ the lattice coordination number.

In model \eqref{eq:ising-Ham}, the nuclear spins are field-polarized near the transition at $A=0$ due to the nuclear Zeeman coupling. Therefore adding a small hyperfine coupling $A\neq 0$ does not change the phase transition qualitatively. However, the transition is generically shifted to larger fields,
as the presence of nuclear spins effectively increases the coordination number of the electronic spins and thus stabilizes symmetry-breaking order.

\begin{figure}[tb]
	\centering
	\includegraphics[width=0.8 \columnwidth]{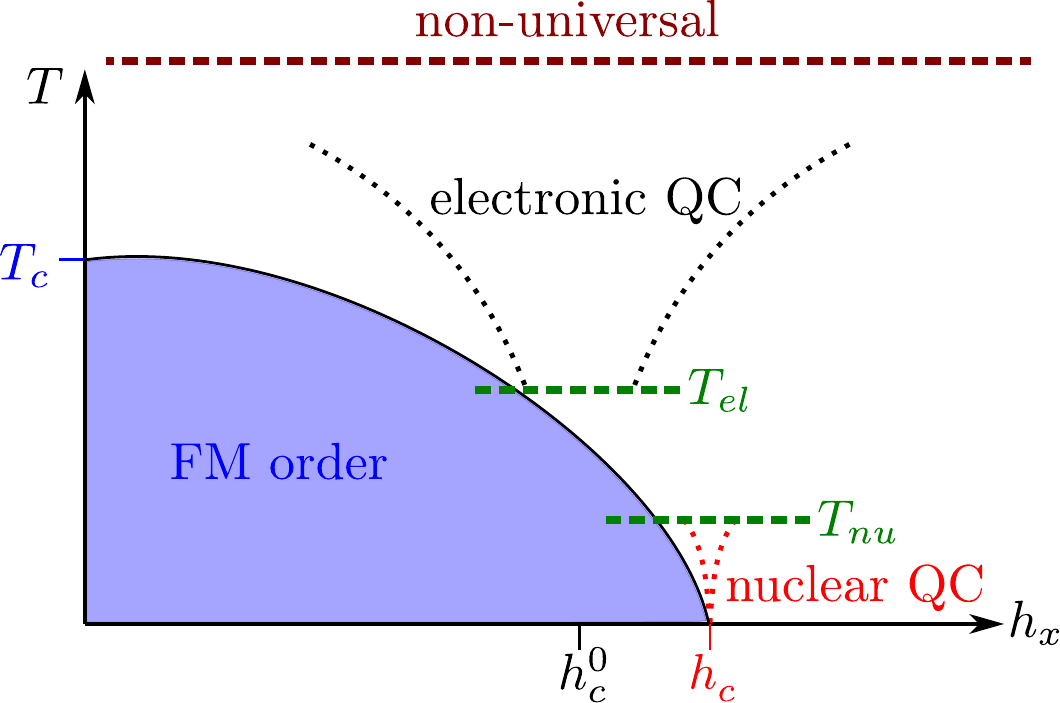}
	\caption{
Schematic temperature--field phase diagram of the transverse-field Ising model \eqref{eq:ising-Ham}: The hyperfine coupling shifts the QPT from $\hcz$ to $\hc$.
While standard electronic quantum criticality (QC) is observable at elevated temperatures, this is cut off below a scale $\Tel$ by hyperfine effects. A novel regime of nuclear quantum criticality emerges below a scale $\Tnu$.
The phase diagram applies to regime II with $A \approx 2d \tilde g_N J $. In contrast, for small $A \ll 2d \tilde g_N J$ distinct regimes of electronic and nuclear criticality do not exist, for details see text.
}
\label{fig:ising-phasediagr}
\end{figure}

Given that Eq.~\eqref{eq:ising-Ham} features two small energy scales, the hyperfine coupling $A$ and the nuclear Zeeman term $\tilde g_N h$, both assumed to be much smaller than the exchange coupling $J$, the discussion of the physics near the quantum phase transition requires to distinguish three parameter regimes depending on the relation between $A$ and $2d \tilde g_N J$:
\begin{itemize}
 \item[I.] $A \ll 2d \tilde g_N J$: In this perturbative limit, the hyperfine coupling induces a weak hybridization between the nuclear mode, which has an energy of order $\tilde g_N h$, and the electronic mode, whose gap closes at the quantum critical point (QCP). At the QCP, the gapless mode is therefore mainly electronic in nature and closely resembles the $A=0$ case. The critical field shifts linearly in $A$ due to the additional contribution $A \langle I_x \rangle$ to the effective field acting on the electrons. Because of its close similarity to the well-known $A=0$ case, regime I will not be discussed in detail here.
 \item[II.] $A \approx 2d \tilde g_N J$: The hyperfine coupling introduces significant hybridization between electronic and nuclear modes. The gapless mode at the QCP carries considerable weight in the nuclear sector, making it meaningful to call it nuclear instead of electronic quantum criticality. Regime II implies substantial modifications of the quantum critical phenomenology, as illustrated in Fig.~\ref{fig:ising-phasediagr}, and it can be realistically found in experiment, requiring hyperfine couplings of order $A/J \approx 10^{-3}$.
 \item[III.] $A > 2d \tilde g_N J$: Larger hyperfine coupling introduces qualitative differences to the $A=0$ case. Most notable is an antiferromagnetic instead of ferromagnetic alignment of the nuclear and electronic spins in the disordered phase above the QCP. Consequently, the boundary between regimes II and III is rather well-defined. A detailed discussion of regime III is relegated to Sec.~\ref{sec:novel}.
\end{itemize}
Where appropriate, the three regimes are marked by different color shading in the subsequent figures.

\subsection{Mean-field theory plus fluctuations}

To study the behavior of model \eqref{eq:ising-Ham} quantitatively, we combine standard mean-field and linear spin-wave theories. We focus on $T=0$ unless otherwise noted.
We first determine the classical reference state by solving the local Hamiltonian
\begin{align}
	\HMF^{\rm TI} = &- 2Jd  \langle S_z \rangle S_z - \vec h (\vec S + \tilde{g}_N \vec I)
	\nonumber \\
	&+ A (\langle \vec S \rangle \cdot\vec I + \vec S \cdot \langle \vec I \rangle - \langle \vec S \rangle \cdot \langle \vec I \rangle )
\end{align}
which results from a mean-field decoupling of both the electronic and hyperfine interactions. The solution can be parameterized by the two angles $\theta = \arctan( \langle S_x \rangle/  \langle S_z \rangle)$ and $\phi = \arctan( -\langle I_x \rangle/  \langle I_z \rangle)$, such that
$\theta = \phi = 0$ in the absence of an external field,
see Fig.~\ref{fig:ising-model}.

Spin-wave theory amounts to an expansion about this classical state, i.e., a product state of individual spins. We rotate the local frame of reference such that all spins point along the (new) $z$-axis, and then introduce Holstein-Primakoff bosons $\hat a_i$, $\hat b_i$ for the electronic and nuclear spins, respectively, for details see Appendix \ref{app:isingcalc}. The resulting bosonic Hamiltonian consists of a constant piece, a bilinear piece and higher-order terms; the linear piece vanishes identically provided that $\theta$ and $\phi$ indeed correspond to the mean-field solution which minimizes $\HMF^{\rm TI}$.
The bilinear part $\mathcal{H}_2^{\rm TI}$ corresponds to linear spin-wave theory; higher-order terms represent boson interactions and will be neglected as they do not lead to qualitative changes of our conclusions.
After Fourier transformation, $\mathcal{H}_2^{\rm TI}$ can be written in matrix form  with $\psi_{\vec k} = \left(a_{\vec k}^{\phantom \dag }, b_{\vec k}^{\phantom \dag }, a_{-\vec k}^\dag, b_{-\vec k}^\dag \right)^T$:
\begin{align}
 \mathcal{H}_2^{\rm TI} =
 &\frac{1}{2} \sum_{\vec k} \left(\psi_{\vec k}\right)^\dag
 \begin{pmatrix}
c_1(\vec k) & c_4	 & c_2(\vec k) & c_3 \\
c_4 & c_5	 & c_3 & 0   \\
c_2(\vec k) & c_3	 & c_1(\vec k) & c_4  \\
c_3 & 0 	 & c_4 & c_5
\end{pmatrix}
\psi_{\vec k}
\end{align}
where we have defined the abbreviations
\begin{align}
 c_1(\vec k) &=  2J d S \cos^2 \theta  + h \sin \theta +  A \alpha I
 	- 2 J d \frac{S}{2} \gamma_{\vec k} \sin^2\theta,
 \nonumber \\
 c_2(\vec k) &= - 2 J d \frac{S}{2} \gamma_{\vec k} \sin^2\theta,
 \nonumber \\
 c_{3,4} &= A \frac{\sqrt{IS}}{2}(\alpha\pm1) ,
  \nonumber \\
 c_5 &= - \tilde g_N h \sin \phi +  A \alpha S
\end{align}
with $\alpha = \cos\theta \cos\phi - \sin\theta \sin\phi$, for details see Appendix~\ref{app:isingcalc}.
Note that only the electronic contributions $c_1$ and $c_2$ display a momentum dependence, as all terms involving nuclear spins are local in space. Without external fields we have $\theta= \phi=0$, and the number-conserving coupling between the boson species vanishes, $c_4=0$, leaving only $S_{\text{tot},z}$-conserving terms ($\vec S_{\text{tot},i} = \vec S_i + \vec I_i$) as dictated by the Heisenberg form of the hyperfine coupling.

The bilinear Hamiltonian $\mathcal{H}_2^{\rm TI}$ can be diagonalized numerically via a bosonic Bogoliubov transformation. We have employed the algorithm presented in the Appendix of Ref.~\onlinecite{Wessel05}. The results of this procedure are described below.

The semiclassical approach to Eq.~\eqref{eq:ising-Ham} is justified as long as the hyperfine coupling is small enough that singlet formation between electronic and nuclear spins plays no role. This is a reasonable approximation for $A\ll J$, typically realized in solids. We note that a more refined mean-field theory will be used in Sec.~\ref{sec:novel} below to study regime III.

\begin{figure}[tb]
	\centering
	\includegraphics[width=\columnwidth]{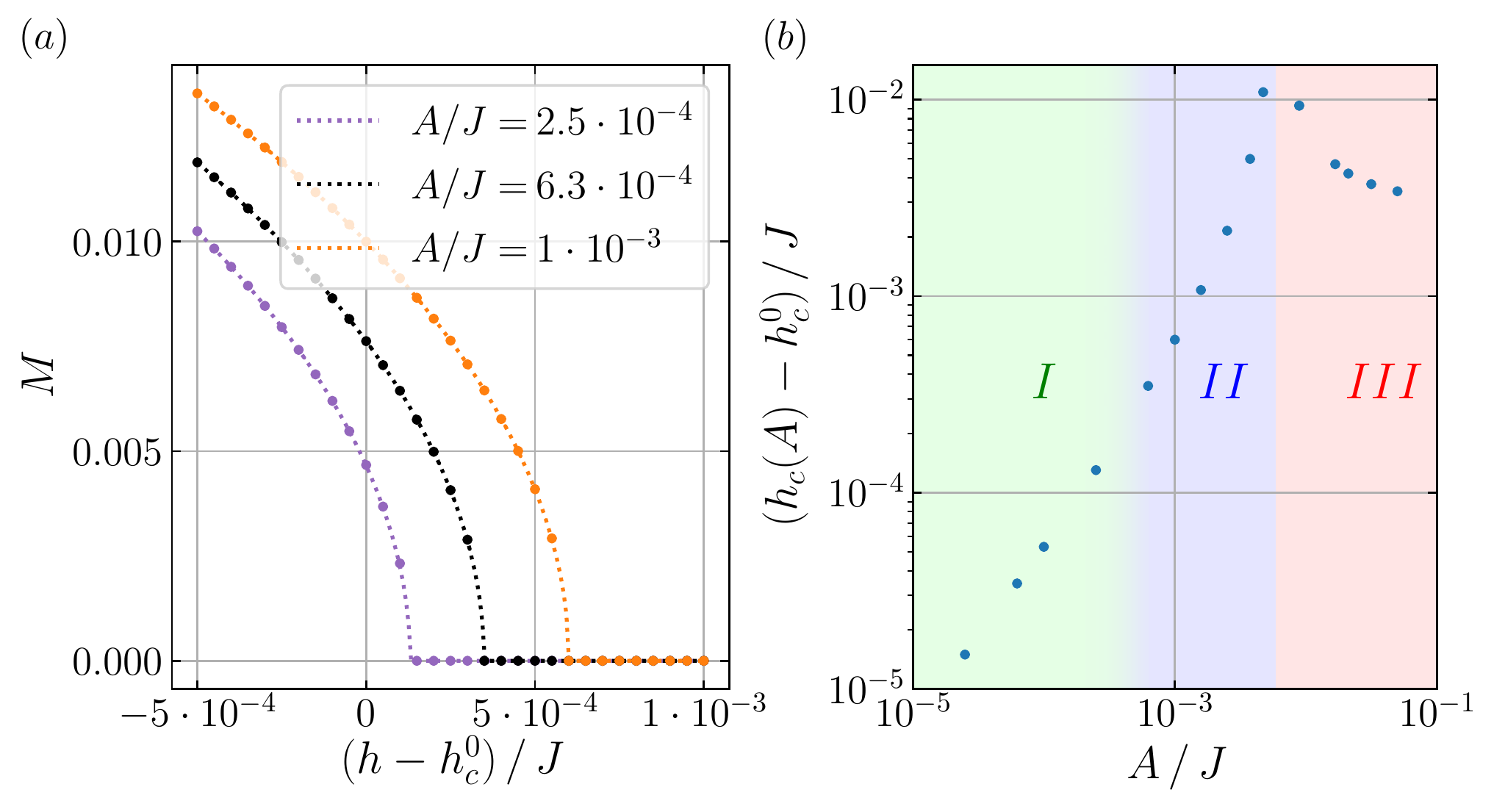}
	\caption{
(a) Order parameter $M$ as function of the transverse field, measured relative to the position $\hcz$ of the QPT at $A=0$, in the Ising model  \eqref{eq:ising-Ham} for different strengths of the hyperfine coupling $A/J$. The data follow a power law $M \propto (h-\hc)^\beta$ (dashed lines), with $\beta=1/2$ as expected for mean-field theory.
(b) Shift of the critical field $\hc$ as function of hyperfine coupling, being linear for small $A$. The shading indicates the three regimes of $A/(2d \tilde g_N J)$ as described in the text; the behavior at large $A$ is discussed in Sec.~\ref{sec:novel}.
}
	\label{fig:ising-MF}
\end{figure}

\subsection{Phase transition and order parameter}

The ferromagnetic phase is characterized by the electronic order parameter $M = \langle S_{iz} \rangle$. Its value including fluctuation corrections can be written as
\begin{align}
 M = \cos \theta \Big( S - \frac{1}{N} \sum_{\vec k} \langle a_{\vec k}^\dag a_{\vec k}^{\phantom \dag} \rangle \Big)
\end{align}
with $N$ the number of lattice sites. The expectation value of the Holstein-Primakoff bosons can be evaluated by inserting the inverse Bogoliubov transformation. As expected, the largest fluctuation corrections appear near the QPT. We also note that cubic terms induce additional $1/S$ corrections to the angle $\theta$. Overall the fluctuation corrections are by a factor $10^{-3}$ smaller than the mean-field value and thus negligible. Results for $M$ for different hyperfine couplings are in Fig.~\ref{fig:ising-MF}.

Upon introducing hyperfine coupling, the field-driven quantum phase transition between the ferromagnetic and paramagnetic phases remains continuous, but is shifted to higher fields, Fig.~\ref{fig:ising-MF}.
The shift $(h_c-h_c^0)$ is initially -- in regime I -- linear in $A$, as expected from the lowering of the effective field acting on the electronic spins $h_{{\rm eff}, x} = h - A \langle I_x \rangle$ due to the antiferromagnetic hyperfine coupling. In regime II, the shift grows faster than linear, while regime III involves qualitative changes, discussed in more detail in Sec.~\ref{sec:novel}.

The hyperfine coupling does not change the universality class of the transition. In our calculation, we observe mean-field exponents which apply above the upper critical dimension, $d\geq d_c^+=3$ (up to logarithmic corrections in $d=3$): The electronic spontaneous magnetization $\langle S^z \rangle$ varies as $(\hc-h)^\beta$ with $\beta=1/2$, see Fig.~\ref{fig:ising-MF}(a).

\begin{figure}[tb]
	\centering
	\includegraphics[width=\columnwidth]{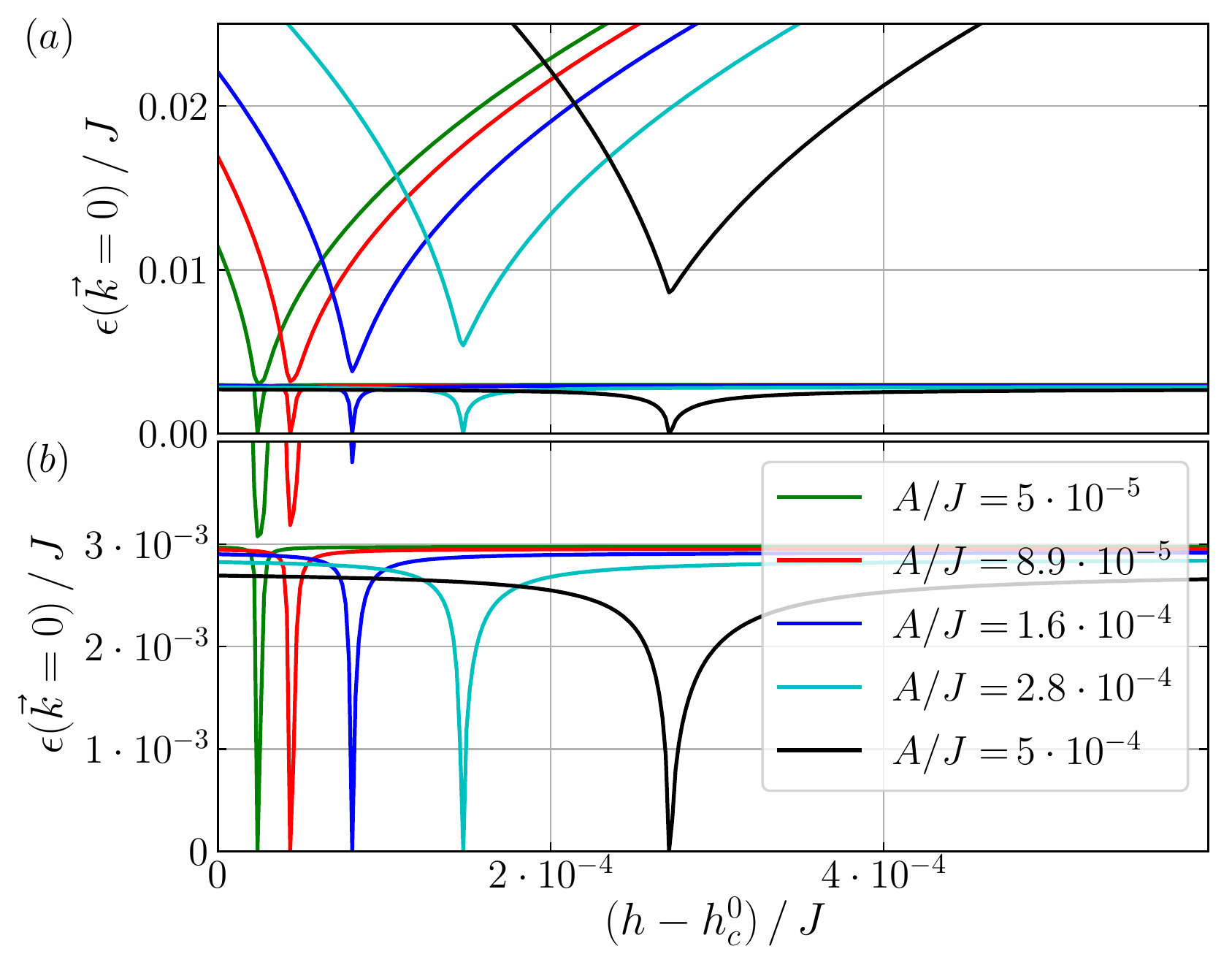}
	\caption{
Mode gaps in the Ising model \eqref{eq:ising-Ham} as function of the transverse field, measured relative to the position $\hcz$ of the QPT at $A=0$, for different values of $A$ in regimes I/II. Panels (a) and (b) show the same data, with (b) zooming into nuclear energy scales.
}
   \label{fig:ising-gap}
\end{figure}

\begin{figure}[tb]
	\centering
	\includegraphics[width=0.8\columnwidth]{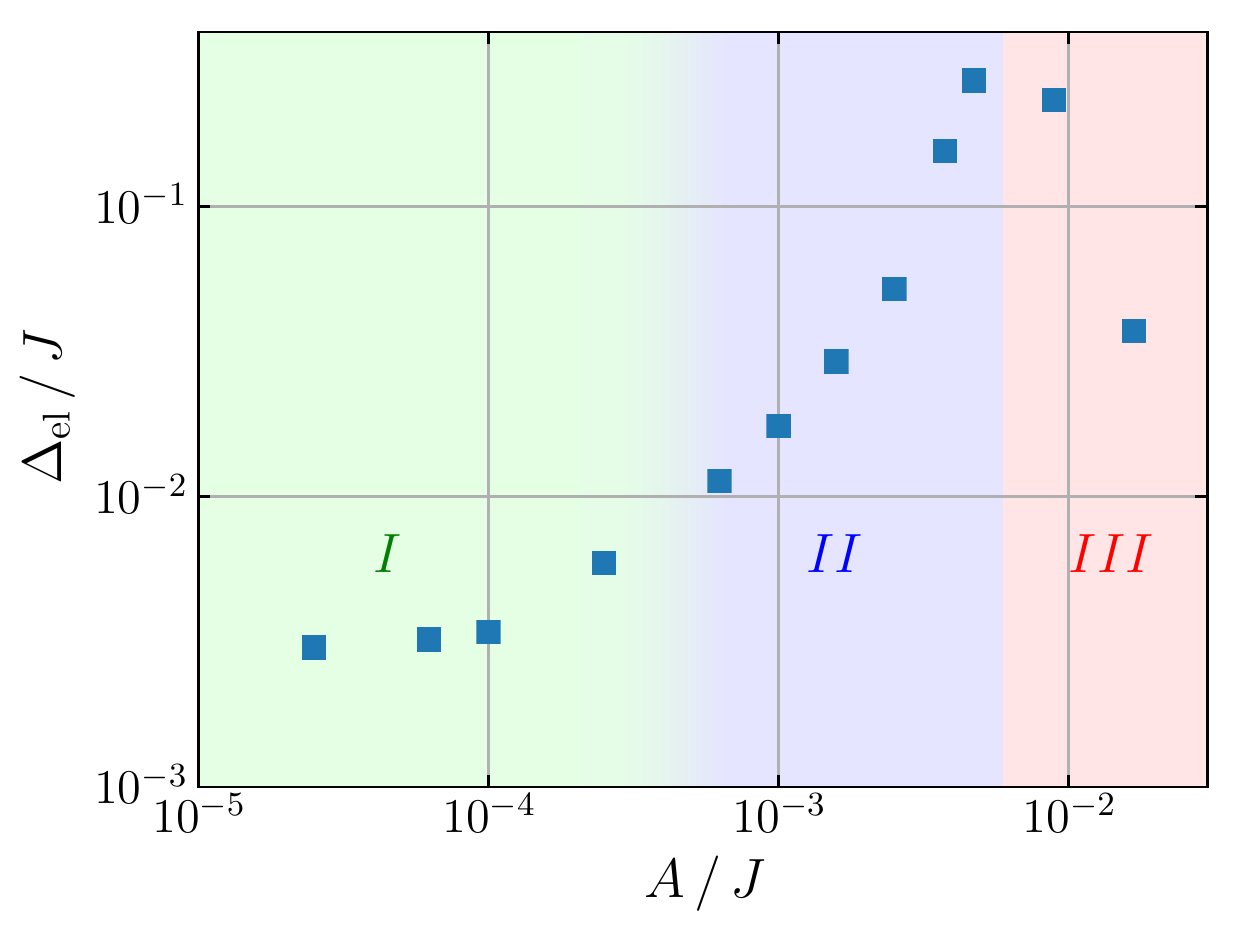}
	\caption{
Gap of the upper (electronic) mode at the QPT, $\Del(\hc)$, as function of hyperfine coupling $A/J$ for the Ising model \eqref{eq:ising-Ham}. The shading indicates the three regimes of $A/(2d \tilde g_N J)$ as described in the text.
}
	\label{fig:ising-elgap}
\end{figure}

\begin{figure}[bt]
	\centering
	\includegraphics[width=\columnwidth]{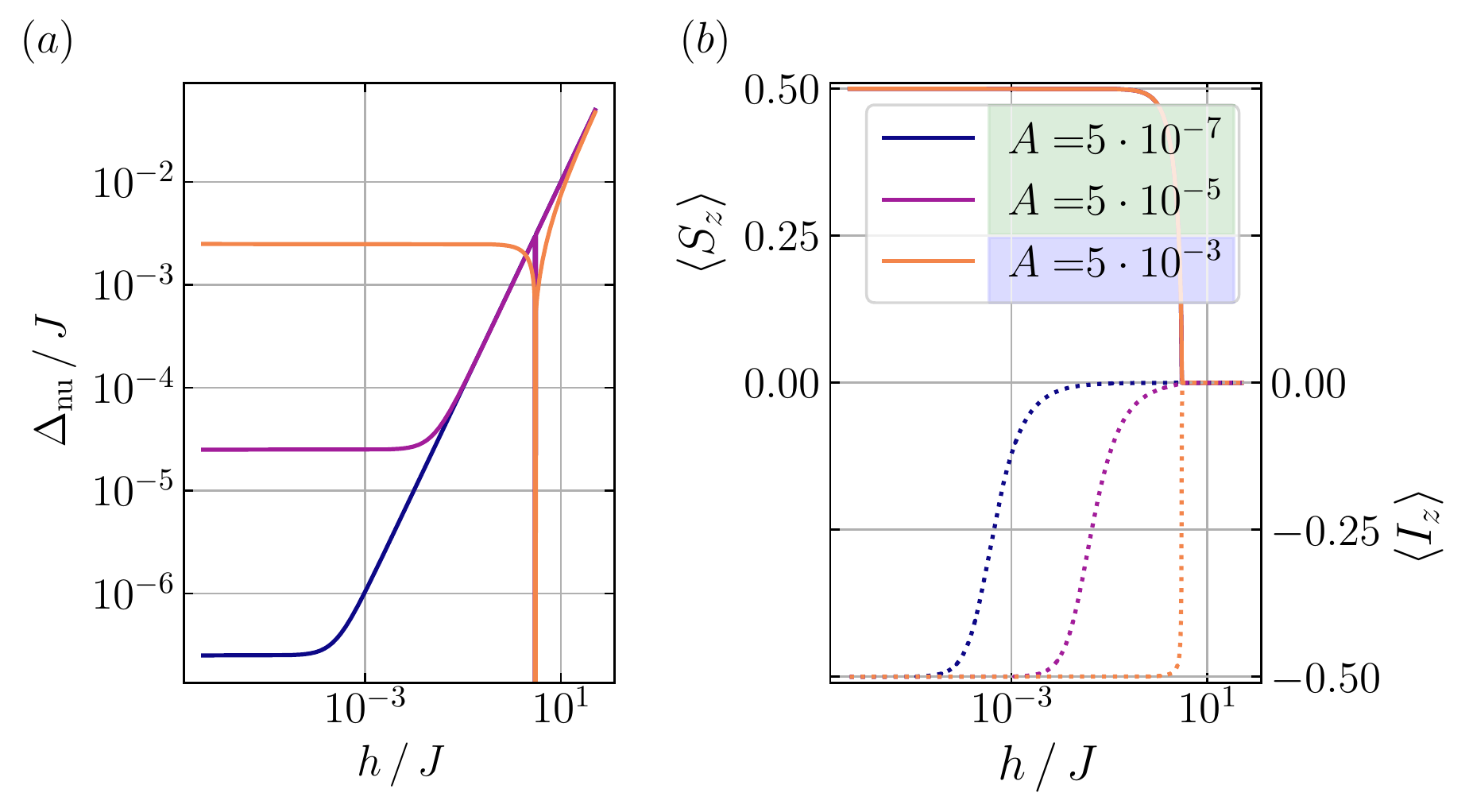}
	\caption{
(a) Energy gap of the lower (nuclear) mode as function of the transverse field $h$ for different hyperfine coupling $A/J$ in the Ising model \eqref{eq:ising-Ham}.
(b) Spontaneous electronic and nuclear magnetizations, $\langle S_z\rangle$ and $\langle I_z\rangle$, as function of $h$ for different $A$.
In both panels, a crossover is seen at $A S \sim \tilde g_N h$ arising from the
competition between hyperfine coupling and nuclear Zeeman term.
}
	\label{fig:ising-nuclmode}
\end{figure}

\begin{figure*}[!bt]
	\centering
	\includegraphics[width=0.95\textwidth]{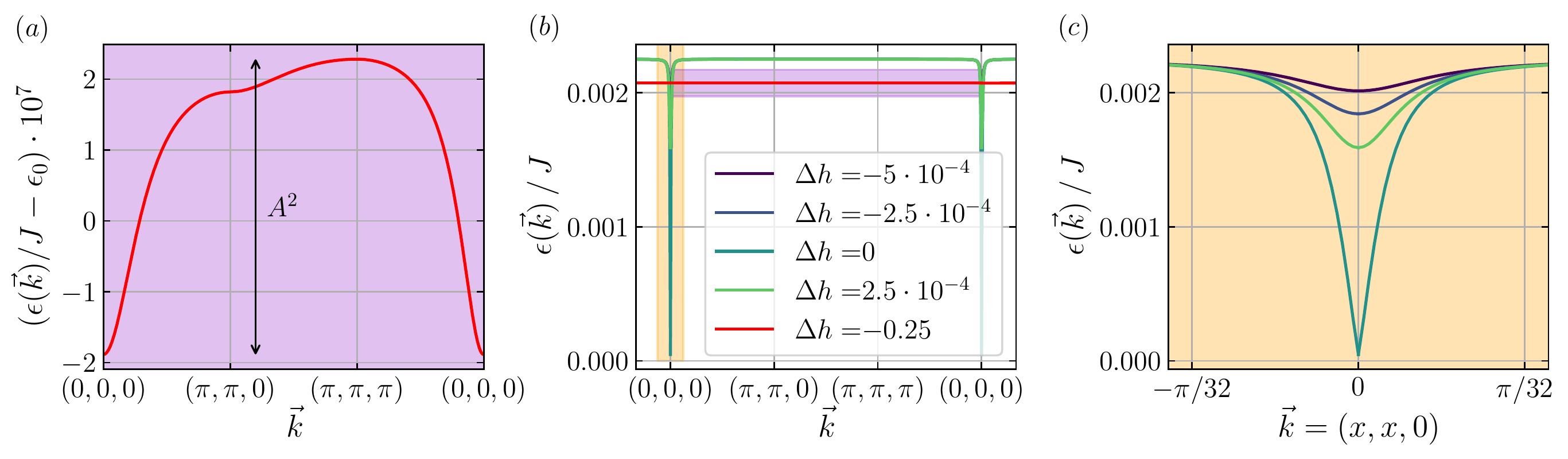}
	\caption{Dispersion of the nuclear mode in model \eqref{eq:ising-Ham} at intermediate $A/J=1.5 \times 10^{-3}$ (regime II) for different field values $\Delta h = h - h_c(A)$ close to the QPT.
Panels (a) and (c) show a zoom of the data in (b) according to the colored boxes (box size not to scale). (a) The bandwidth away from the QPT scales as $A^2/J$; the energy offset here is $\epsilon_0=0.00207245$. (c) At the QPT the gap closes linearly, as expected for a transition with a non-conserved order parameter and $z=1$, while away from criticality the band minimum is quadratic.
}
	\label{fig:ising-nucldisp}
\end{figure*}

\subsection{Excitation spectrum}
\label{sec:isingspec}

The excitation spectrum consists of two modes: For $A=0$ these are a dispersionless nuclear mode and an electronic mode which disperses for $h\neq 0$ and becomes gapless at $h=\hcz$. For any non-zero $A$, both modes are mixtures of electronic and nuclear excitations. We label the mode energies $\eps_{1,2}(\vec k)$ with $\eps_1<\eps_2$, such that the lower (upper) mode is primarily of nuclear (electronic) character, except in regime I in the immediate vicinity of $\vec k = 0$ and $h=\hc$.
Both modes have their dispersion minimum at $\vec k=0$, and we can define two energy gaps $\Dnu = \epsilon_1(\vec k\!=\!0)$ and $\Del = \epsilon_2(\vec k\!=\!0)$.

For non-zero $A$, the quasiparticle spectrum is gapped everywhere except at the quantum phase transition where the lower mode becomes soft, i.e., $\Dnu=0$ at $h=\hc$. This is shown in Fig.~\ref{fig:ising-gap}(b), which also illustrates the shift of $\hc$ with increasing $A$. Near the critical field we have $\Dnu \propto |h-\hc|^{\nu z}$, with $\nu$ and $z$ being the correlation-length and dynamic critical exponents, here $z=1$ and $\nu=1/2$ at mean-field level.
In contrast, the gap of the upper (electronic) mode, $\Del$, is always finite, with a cusp-like minimum at $\hc$. This incomplete electronic mode softening, seen in Fig.~\ref{fig:ising-gap}(a), is a key characteristic of hyperfine interactions at QPTs. The gap $\Del$ is shown in Fig.~\ref{fig:ising-elgap}: In regime I where $A \ll 2d \tilde g_N J$ this gap is determined by the nuclear Zeeman energy and acquires corrections linear in $A$ with increasing $A$. In regime II these contributions dominate and get amplified by the faster-than-linear-in-$A$ shift of the critical field, Fig.~\ref{fig:ising-MF}. As a result, $\Del(\hc)$ defines an energy scale $k_B \Tel$ at criticality which is parametrically larger than $A$ and below which no electronic excitations exist.
In regime III the gap decreases, which is plausible considering the decrease of the critical field.

The behavior of the nuclear mode in regimes I and II is further illustrated in Figs.~\ref{fig:ising-nuclmode} and \ref{fig:ising-nucldisp}. Fig.~\ref{fig:ising-nuclmode}(a) shows that the nuclear-mode energy, and thus its gap away from criticality, is determined by the hyperfine coupling for $\tilde g_N h \ll A$ and by the nuclear Zeeman energy for $\tilde g_N h \gg A$. Concomitantly, the nuclear spins are antialigned with the electronic moments for $\tilde g_N h \ll A$, while they are field-polarized for $\tilde g_N h \gg A$, Fig.~\ref{fig:ising-nuclmode}(b). As a result, nuclear spins hardly participate in the magnetic quantum phase transition in regime I, where they are essentially field-polarized at $\hc$, but do so in regime II, where they stay ordered up to $\hc$.
Away from criticality, the nuclear-mode dispersion is cosine-like, with a small bandwidth of order $A^2/\Del$ due to the electron-mediated interaction between nuclear spins, Fig.~\ref{fig:ising-nucldisp}(a). Close to criticality, however, this interaction gets increasingly long-ranged, such that an anomalous dispersion shape emerges, where $\eps_1(\vec k)$ is essentially flat except for a small vicinity of $\vec k=0$ and the critical-mode velocity at $\hc$ (in units of the lattice constant) is much larger than the bandwidth, Fig.~\ref{fig:ising-nucldisp}(b,c).

\begin{figure}[tb]
	\centering
	\includegraphics[width=\columnwidth]{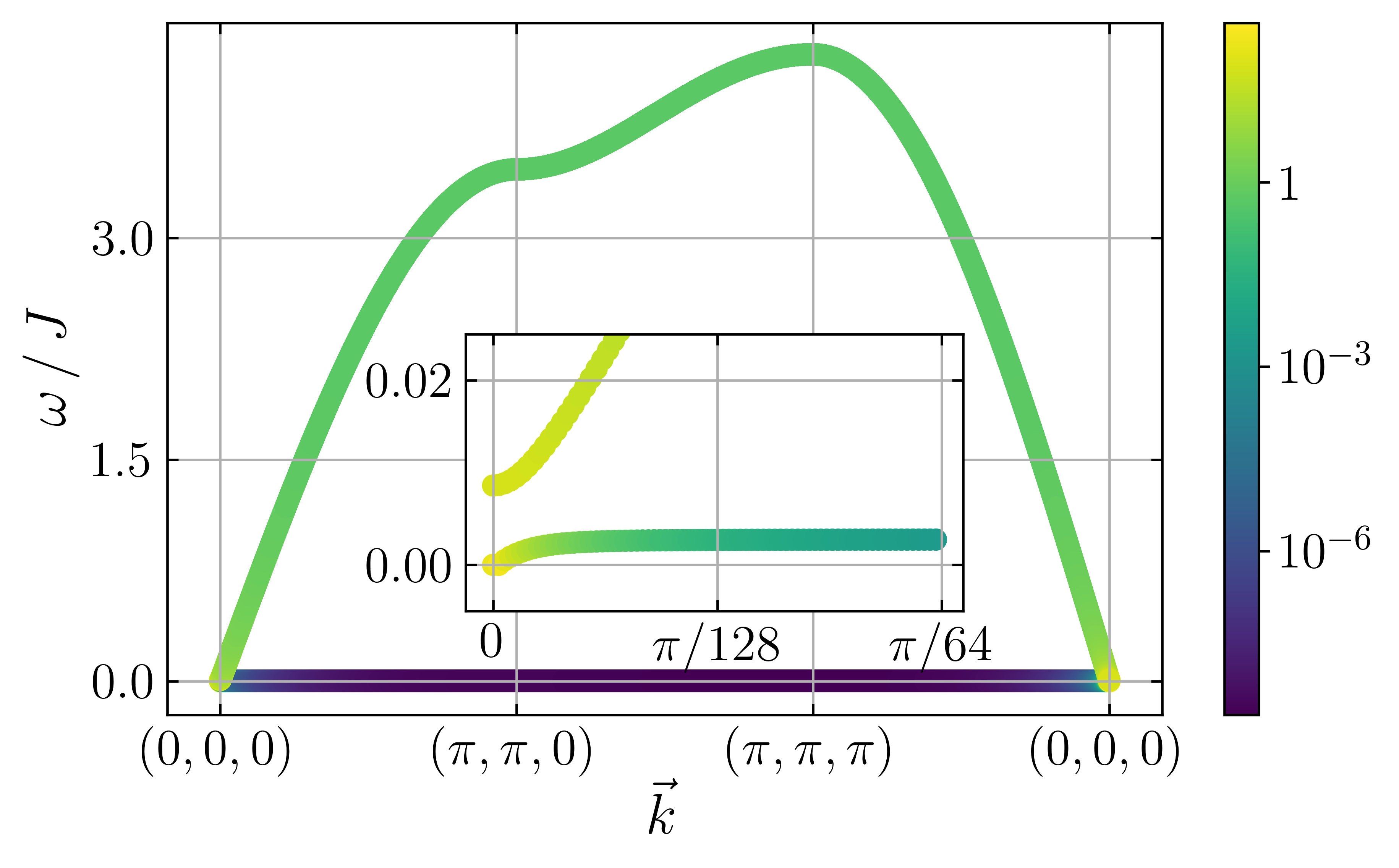}
	\caption{
Imaginary part of the transverse electronic dynamical susceptibility ${\rm Tr}\,\chi_{\alpha\beta}''(\vec k,\w)$ for the Ising model \eqref{eq:ising-Ham}, as function of $\vec k$ and $\w$ and calculated at the QPT with intermediate hyperfine coupling $A/J=5 \times 10^{-4}$. The color code shows the mode weight on a logarithmic scale; the linewidth is artificial.
The inset illustrates the mode hybridization at small energies and wavevectors.
}
\label{fig:ising-susc}
\end{figure}

The eigenmodes of $\mathcal{H}_2^{\rm TI}$ can be used to calculate the electronic dynamical spin susceptibility, as measured by inelastic neutron scattering:
\begin{align}
	\chi_{\alpha \beta}(\vec k, \omega) = -i \int_{-\infty}^\infty dt \, e^{i \omega t} \langle \mathcal{T}_t S_\alpha(\vec k, t) S_\beta(-\vec k, 0)\rangle
\label{eq:chi}
\end{align}
with $\alpha,\beta \in \{x,y,z\}$. We restrict ourselves to the single-mode approximation where each spin operator translates into one Holstein-Primakoff boson. The resulting imaginary part, $\chi''(\vec k, \omega)$, displays sharp peaks at the quasiparticle dispersion $\w = \epsilon_{1,2}(\vec k)$. Higher-order terms will be neglected, as they contribute to multi-particle continua only.

Results are shown in Fig.~\ref{fig:ising-susc} for the quantum critical point, $h=\hc$. The intensity distribution nicely illustrates the hybridization between electronic and nuclear modes. Further, the intensity of the nuclear mode diverges at criticality due to the gap closing, technically due to the properties of the Bogoliubov transformation. We recall that a sharp critical mode is a feature of mean-field theory and is only expected at or above the upper critical dimension; otherwise a power law of the type $\chi_{\alpha\beta}(\vec k, \omega) \propto (\eps^2(k)-\w^2)^{-1+\eta/2}$ occurs, with $\eta$ an anomalous exponent. \cite{ssbook}

\subsection{Thermodynamics}

While the previous results have been restricted to $T=0$, we now use these insights to discuss the finite-temperature behavior near the QPT, resulting in the phase diagram shown in Fig.~\ref{fig:ising-phasediagr}. While regime I of small hyperfine coupling displays little difference to the $A=0$ (i.e. purely electronic) QPT scenario, regimes II and III feature (i) a QPT which is significantly shifted to higher fields (compared to the $A=0$ case), and (ii) distinct temperature regimes of electronic and nuclear quantum criticality.

Generically, a quantum critical regime displays power-law behavior of thermodynamic observables as function of temperature. For instance, the specific heat follows the scaling prediction $C(T)\propto T^{d/z}$ below the upper critical dimension $d_c^+$; for $d=d_c^+$ logarithmic corrections to this power law occur. For the transverse-field Ising QPT we recall $z=1$ and $d_c^+=3$.

As argued above, sizeable hyperfine coupling opens an energy window $\tilde g_N h \equiv k_B \Tnu \ll k_B T \ll k_B \Tel$ between the nuclear and electronic modes at criticality -- this applies to parameter regimes II and III. As a result, distinct and separate electronic and nuclear quantum critical regimes exist, as shown in Fig.~\ref{fig:ising-phasediagr}.
Both quantum critical regimes will separately display specific-heat power laws, with the prefactor of the low-$T$ nuclear regime being significantly larger than that of the electronic regime due to the smaller nuclear energy scales. From the critical mode velocities in our mean-field-based calculation (at $d=d_c^+$) we may deduce the ratio of specific-heat prefactors, $\kappa_{\rm el, nu} = C_{\rm el, nu}/T^3$ and obtain, e.g., $\kappa_{\rm nu}/\kappa_{\rm el} \approx 5 \times 10^{3}$ at $A/J=1.5\times 10^{-3}$. In the window $\Tnu \ll T \ll \Tel$ the specific heat will be small, thus facilitating the crossover between both critical power laws.

\subsection{LiHoF$_4$} \label{sec:LHF}

{\lhf} is considered a prime example for transverse-field Ising quantum criticality. It features a sizeable hyperfine coupling to nuclear spins $I=7/2$ of the magnetic Ho$^{3+}$ ions. Indeed, the hyperfine coupling has been argued to be responsible for a significant increase of the critical field.\cite{bitko96, Chakraborty04, Tabei08, McKenzie18}
The incomplete softening of the electronic mode at the QCP, explained in Sec.~\ref{sec:isingspec} above, has been observed via inelastic neutron scattering \cite{Ronnow05} and microwave spectroscopy \cite{Kovacevic16} and also theoretically investigated in Ref.~\onlinecite{McKenzie18}.

While the qualitative phenomenology of {\lhf} agrees well with our analysis, a direct quantitative comparison with the calculations presented above is, however, not straightforward, because the Ho$^{3+}$ ions are characterized by a non-Kramers ground-state doublet of the crystalline electric field. This leads to large variations of the electronic magnetic moment $J=(J_x^2+J_y^2+J_z^2)^{1/2}$ as function of the external field:\cite{ourLHFpaper} The moment $J$ is larger in the ordered phase than in the disordered phase close to the QCP. As a result, the hyperfine coupling amplifies Ising order more strongly than in the case of Kramers moments, as employed in the calculation of this section.


\section{Smeared transitions: Coupled-dimer magnets under pressure}
\label{sec:dimers}

\begin{figure}[tb]
	\centering
	\includegraphics[width=0.6 \columnwidth]{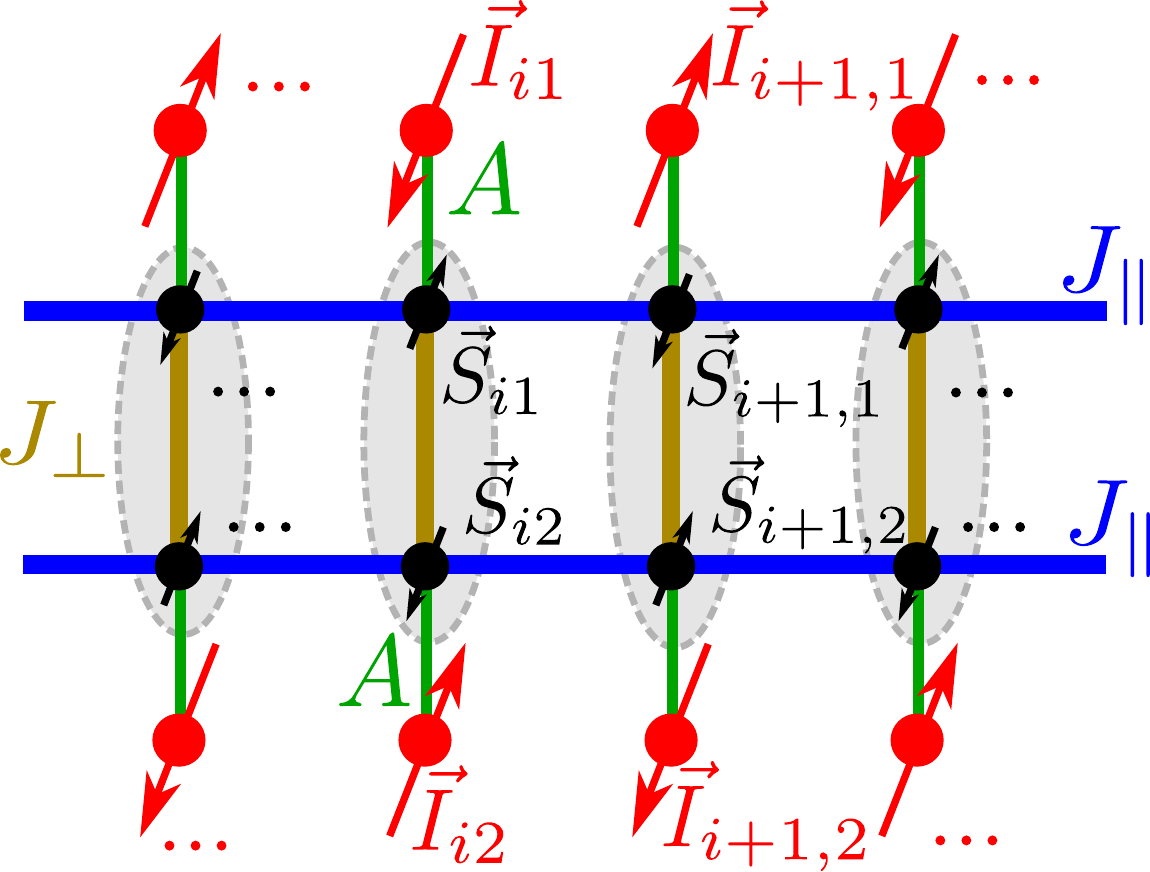}
	\caption{
    Coupled-dimer Heisenberg model with nuclear spins (schematic): electronic spins (black arrows)  are pairwise coupled by the intradimer coupling $J_\perp$ (yellow thick lines), which drives singlet formation (grey ellipses). Spins in neighboring pairs are coupled by the interdimer coupling $J_\parallel$ (blue thick lines). The hyperfine coupling $A$ (green lines) links electronic and nuclear spins (red arrows).
}
	\label{fig:dimers-model}
\end{figure}

We now turn to quantum transitions which are destroyed, i.e., smeared, by the presence of nuclear spins. As explained in Sec.~\ref{sec:symm}, this happens for order--disorder transitions where the hyperfine coupling induces weak magnetic order in the otherwise paramagnetic (and time-reversal-symmetric) electronic phase.
As an example we consider a coupled-dimer magnet which can be driven from a singlet dimer phase to an antiferromagnetic phase by varying the ratio of exchange couplings, experimentally done, e.g., by pressure. The Hamiltonian, augmented by nuclear spins, reads:
\begin{align}
  \mathcal{H}^{\rm CD} &= J_\perp \sum_i \vec S_{i1} \cdot  \vec S_{i2} + J_\parallel \sum_{\langle ij \rangle, m} \vec S_{im} \cdot  \vec S_{jm}
  \nonumber \\  &
  ~~+ A \sum_{i,m} \vec S_{im} \cdot  \vec I_{im}
  \label{eq:dimers-Ham}
\end{align}
where $i,j$ denote dimer sites, and $m \in \{1,2\}$ labels the spins in each dimer. The intradimer ($J_\perp$), interdimer ($J_\parallel$), and hyperfine ($A$) couplings are all assumed to be of Heisenberg type and antiferromagnetic ($\geq 0$), such that the model displays a global SU(2) symmetry of combined rotations of electronic and nuclear spins. A sketch of the system is shown in Fig.~\ref{fig:dimers-model}.
For concreteness we consider a cubic lattice of dimers. It is useful to introduce $q= d J_\parallel/J_\perp$, with $2d=6$ the lattice coordination number, as a dimensionless measure of the interdimer coupling strength. All numerical results in the following will be presented for $S=1/2$ and $I=1/2$.

\subsection{General considerations}

In the absence of nuclear spins the coupled-dimer magnet displays a QPT between a gapped quantum paramagnetic dimer phase, realized for $q<\qcz$, and an antiferromagnetic (AF) phase for $q>\qcz$.\cite{Sachdev90, Chubukov95, Vojta99, Wang06, Giamarchi08} For a cubic lattice of dimers the antiferromagnet displays collinear order at wavevector ${\vec Q}= (\pi,\pi,\pi)$.
%

Including the nuclear spins adds a manifold of states which is massively degenerate at $A=0$ and whose fate is thus determined by hyperfine coupling. Treating $A$ perturbatively, the nuclear spins align with the mean field coming from the electronic bulk order in the AF phase. However, in the dimer phase such a mean-field is absent, and the leading effect is an indirect interaction between the nuclear spins mediated by the electronic paramagnet. This interaction scales as $A^2/\Delta_0$ where $\Delta_0$ is the spin gap of the electronic subsystem in the absence of nuclear spins.
For unfrustrated lattices, this effective interaction oscillates in sign with distance and induces low-temperature AF order of the nuclear spins at the same wavevector $\vec Q$ where the electronic system tends to order. By proximity, the nuclear-spin order also induces weak AF order on the electronic spins via a staggered mean field.

\begin{figure}[bt]
	\centering
	\includegraphics[width=0.8\columnwidth]{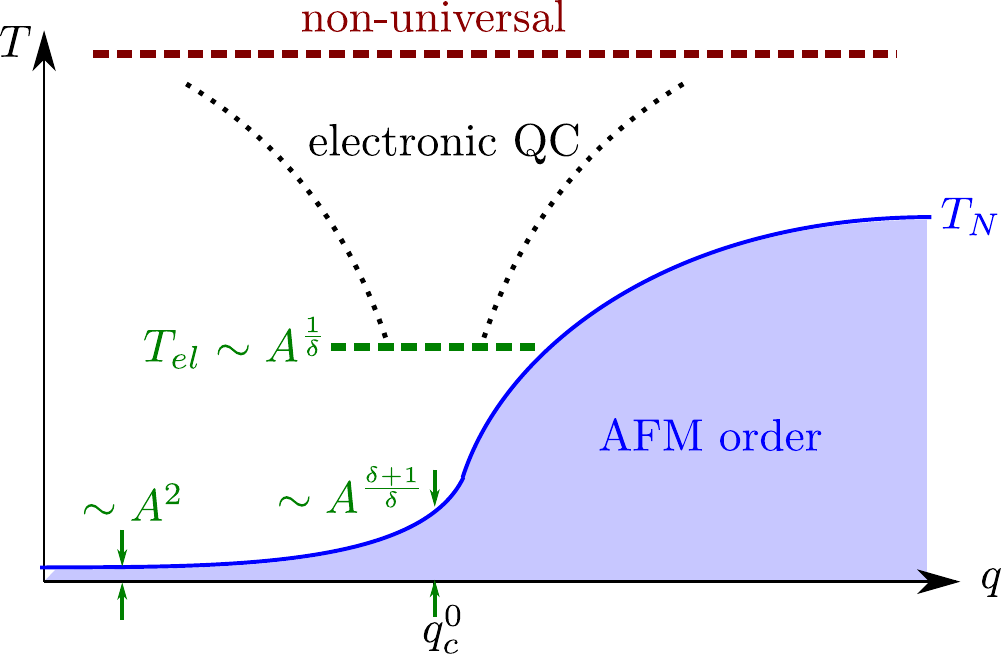}
	\caption{
Schematic phase diagram of the coupled-dimer model \eqref{eq:dimers-Ham}: In the presence of a finite hyperfine coupling $A$ the disordered phase is replaced by a weak antiferromagnet, turning the quantum phase transition at $\qcz$ into a crossover. The electronic quantum criticality is cut off below a scale $\Tel$ set by the triplon gap. The N\'eel temperature $T_N$ for $q\leq\qcz$ is determined by $A$, for details see text.
}
	\label{fig:dimers-phasediagram}
\end{figure}

Consequently, the electronic dimer phase at $q<\qcz$ is unstable: Any small hyperfine coupling turns the system into a weak SU(2)-breaking antiferromagnet. Therefore the QPT is smeared into a crossover, as illustrated in the phase diagram in Fig.~\ref{fig:dimers-phasediagram}. Signatures of quantum criticality are observable only above a hyperfine-induced temperature (or energy) scale $\Tel$. As will become clear below, all hyperfine-induced energy and temperature scales can be related to powers of $A$, because the dimer model contains no additional small energy scale, in contrast to the Ising model in Sec.~\ref{sec:ising} with its nuclear Zeeman term.

\subsection{Mean-field theory plus fluctuations}

To faithfully describe the behavior of the model \eqref{eq:dimers-Ham}, we resort to a combination of bond-operator and spin-wave theories.
Bond-operator theory \cite{Sachdev90} is a method developed to describe the excitation spectrum of coupled-dimer magnets and rests on using singlet and triplet states as a basis for the local Hilbert space of each dimer, in our case made up of the spins $\vec S_{i1}$ and $\vec S_{i2}$. The bond-operator description expands around a dimer product state and treats its gaussian fluctuations as bosonic particles. While originally used for the paramagnetic phase only, with a singlet reference state and triplon excitations, bond-operator theory has later been extended to magnetically ordered phases by a suitable modification of the reference state.\cite{Sommer01,Joshi15b} Refs.~\onlinecite{Joshi15a,Joshi15b} showed that bond-operator theory can be  understood and controlled in terms of a $1/d$ expansion, and we will use it here for the electronic spins in the model \eqref{eq:dimers-Ham}.

As argued above, the model Eq.~\eqref{eq:dimers-Ham} shows weak magnetic order even for small $q$ once the hyperfine coupling is non-zero. This magnetic order is driven and carried by the nuclear spins. Hence the nuclear spins can be assumed to be reasonable well ordered for any value of $q$ which justifies the use of spin-wave theory for the nuclear spins.\cite{Kittel,Stancil}

\subsubsection{Product state}

The model \eqref{eq:dimers-Ham} features four spins per crystallographic unit cell. Given the presence of magnetic order at low $T$ for all $q$, we shall employ a product reference state of the form
\begin{equation}
|\psi_0\rangle = \prod_i |\tilde s;\downarrow,\uparrow\rangle_i
\label{eq:ps}
\end{equation}
where $|\tilde s;I_1^z,I_2^z\rangle_i$ is a state on site $i$ where $\tilde s$ refers to an entangled state of the two electronic spins $S_{im}$ whereas $I_{1,2}^z$ correspond to a classically aligned state of the two nuclear spins, see Fig.~\ref{fig:dimers-model}.
Following Refs.~\onlinecite{Sommer01,Vojta13,Joshi15b} we choose
\begin{equation}
|\tilde s\rangle_i = \frac{|s \rangle_i + \lambda_i |t_z\rangle_i}{\sqrt{1+\lambda_i^2}}
\label{eq:lambdadef}
\end{equation}
where \mbox{$|s \rangle = (|\uparrow\downarrow\rangle - |\downarrow\uparrow\rangle)/\sqrt{2}$} and \mbox{$|t_z \rangle = (|\uparrow\downarrow\rangle + |\downarrow\uparrow\rangle)/\sqrt{2}$} are the intra-dimer singlet and $z$-triplet states, respectively.
The $\lambda_i$ parameterize a rotation in the local SU(4) space of dimer states and are chosen as $\lambda_i = e^{i \vec Q \vec r_i} \lambda$ in order to imprint staggered spin order along the $z$-direction. The variational parameter $\lambda$ enables an interpolation between a singlet paramagnet of electronic spins ($\lambda=0$) and a fully polarized antiferromagnet ($\lambda=\pm1$).
We note that the product state \eqref{eq:ps} neglects local entanglement between electronic and nuclear spins, which is justified provided that the hyperfine coupling $A$ is small, $A\ll J_\perp, 2d J_\parallel$.

\begin{figure}[tb]
	\centering
	\includegraphics[width=\columnwidth]{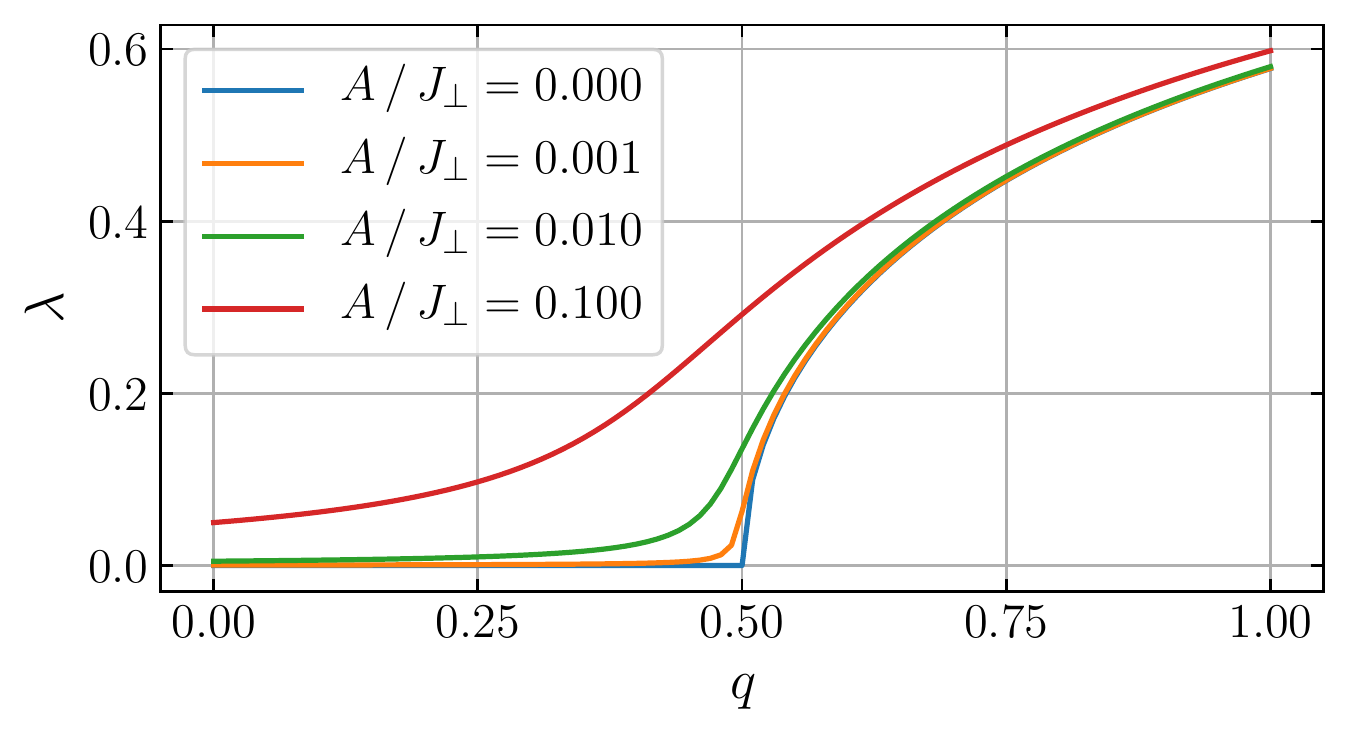}
	\caption{
Rotation parameter $\lambda$ \eqref{eq:lambdadef} as function of inter-dimer coupling $q$ for different values of the hyperfine coupling $A$ in model \eqref{eq:dimers-Ham}. Finite $A$ smears the transition from the paramagnetic phase ($\lambda =0$) to the AFM phase ($\lambda \neq 0$).
}
\label{fig:dimers-lambdaOfq}
\end{figure}

The optimal value for $\lambda$ as function of the model parameters $q$ and $A$ is determined by minimizing the energy of the product state, $\Eps = \langle \psi_0 | \mathcal{H}^{\rm CD} | \psi_0 \rangle$. We obtain
\begin{align}
	\frac{\Eps}{N J_\perp} =
	 - \frac{2 \lambda}{1+\lambda^2} \frac{A I+ \hst}{J_\perp}
	 + \frac{(-3 + \lambda^2)}{4(1+\lambda^2)}
	 - \frac{2 q \lambda^2}{(1+\lambda^2)^2}
\end{align}
where $\hst$ represents an additional staggered field along $\hat z$ applied to the electronic spins, included for later convenience. The expression shows that the hyperfine coupling acts like a staggered field as well.
Minimizing $\Eps$ yields results for $\lambda(q,A)$ as shown in Fig.~\ref{fig:dimers-lambdaOfq}. Without hyperfine coupling, $A=0$, the rotation parameter vanishes for all $q<\qcz$, i.e., in the dimer phase of the electronic model, while it is finite for $q>\qcz$, corresponding to AF order. On this level of approximation, the transition occurs at $\qcz=1/2$.
At non-zero hyperfine coupling, $A>0$, we find $\lambda \neq 0$ for all values of the intra-dimer coupling. This underlines that any finite hyperfine coupling destroys the QPT by destabilizing the paramagnetic ground state. In the low-temperature limit, only a crossover from weak to strong antiferromagnetism remains.

\subsubsection{Gaussian fluctuations}

Fluctuations on top of the product state \eqref{eq:ps} can be captured in a bosonic formalism.
For the electronic spins we define generalized triplon operators\cite{Sommer01,Vojta13,Joshi15b} $\tilde t_{i\alpha}$, $\alpha=x,y,z$. These create local excitations on site $i$, \mbox{$\tilde t_{i\alpha}^\dag |\tilde s \rangle_i = |\tilde t_\alpha \rangle_i$}, where the excited states are
\begin{align}
	|\tilde t_x \rangle_i &= |t_x \rangle_i = (-|\uparrow\uparrow \rangle_i + |\downarrow\downarrow\rangle_i)/\sqrt{2},
	\nonumber \\
	|\tilde t_y \rangle_i &= |t_y \rangle_i = i(|\uparrow\uparrow \rangle_i + |\downarrow\downarrow\rangle_i)/\sqrt{2},
	\nonumber \\
	|\tilde t_z \rangle_i &= ( - \lambda_i |s\rangle_i + |t_z\rangle_i)/\sqrt{1+\lambda^2}.
\end{align}
The physical dimer Hilbert space corresponds to the hard-core condition $\sum_\alpha \tilde t_{i\alpha}^\dag \tilde t_{i\alpha} \leq 1$.
Excitations of the nuclear spins are captured by Holstein-Primakoff bosons, $a_i$ and $b_i$ for the two nuclear spins per lattice site, as in standard spin-wave theory.\cite{Kittel,Stancil}

Expressing all spin operators in terms of the five bosons $\{\tilde t_{x},\tilde t_{y},\tilde t_{z},a,b\}$, inserting this into the Hamiltonian \eqref{eq:dimers-Ham}, and implementing the triplon hard-core constraint via projectors\cite{Joshi15a,Joshi15b} yields an expression of the form $\mathcal{H}_0 + \mathcal{H}_1+ \mathcal{H}_2+ \ldots$ where $\mathcal{H}_n$ contains $n$ bosonic operators, for details of the calculation see Appendix \ref{app:dimerscalc}.
$\mathcal{H}_0$ is a constant which equals $\Eps$. The coefficients of the linear piece $\mathcal{H}_1$ vanish identically provided that $\lambda$ minimizes $\Eps$. In the following, we restrict ourselves to an analysis of the bilinear piece $\mathcal{H}_2$ which describes Gaussian fluctuations. Higher-order terms corresponding to boson interactions will be neglected for simplicity; as a result all critical exponents of the electronic problem take mean-field values.\cite{Joshi15b}

For the collinear reference state \eqref{eq:ps} the excitations created by $\{\tilde t_x,\tilde t_y,a,b\}$ correspond to transverse fluctuations, and we observe that in $\mathcal{H}_2$ the $x$ ($y$) triplons couple only to $c_i = [a_i - b_i]/\sqrt{2}$ ($d_i = e^{i \vec Q \vec r_i} [a_i + b_i]/\sqrt{2}$), respectively. In contrast, the excitations created by $\tilde t_z$ correspond to longitudinal fluctuations which do not couple to the transverse ones in $\mathcal{H}_2$.
After Fourier transformation, the degrees of freedom can therefore be grouped as
$\psi^{(x)}_{\vec k} = (\tilde t_{\vec k x}, c_{\vec k }, \tilde t_{-\vec k x}^\dag, c_{-\vec k }^\dag)^T$,
$\psi^{(y)}_{\vec k} = (\tilde t_{\vec k y}, d_{\vec k}, \tilde t_{-\vec k y}^\dag,  d_{-\vec k}^\dag)^T$ and $\psi^{(z)}_{\vec k} = \left( \tilde t_{\vec k z}^{\phantom \dag}, \tilde t_{-\vec k z}^\dag \right)^T$.
The resulting bilinear Hamiltonian reads
\begin{align}
 \mathcal{H}_2^{\rm CD} =
 &\frac{J_\perp}{2} \sum_{\alpha \in \{ x,y,z\} } \sum_{\vec k} \left(\psi^{(\alpha)}_{\vec k}\right)^\dag
	M^{(\alpha)}(\vec k)
\psi^{(\alpha)}_{\vec k}
\label{eq:h2dim}
\end{align}
with the Hamiltonian matrices
\begin{align}
	M^{(x)}(\vec k) =
	\begin{pmatrix}
c_1(\vec k) & c_4 & c_2(\vec k) & c_3 \\
c_4 & c_5	& c_3 & 0 &  \\
c_2(\vec k) & c_3 & c_1(\vec k) & c_4  \\
c_3 & 0     & c_4 & c_5
\end{pmatrix},
\end{align}
\begin{align}
	M^{(y)}(\vec k) =
 \begin{pmatrix}
c_1(\vec k) & -ic_4  & c_2(\vec k) & ic_3 \\
ic_4 & c_5	 & ic_3 & 0 &  \\
c_2(\vec k) & -ic_3 & c_1(\vec k) & ic_4  \\
-i c_3 & 0  & -ic_4 & c_5
\end{pmatrix},
\end{align}
and
\begin{align}
	M^{(z)}(\vec k) =
 \begin{pmatrix}
c_6(\vec k) & c_7(\vec k)  \\
c_7(\vec k) & c_6(\vec k)	
\end{pmatrix}.
\end{align}
The coefficients appearing in the Hamiltonian matrices are
\begin{align}
  c_1(\vec k) =& \frac{1}{1+\lambda^2} \left[ - 2 \frac{A I+\hst}{J_\perp} \lambda + 1  + 4 \frac{q \lambda^2}{1+\lambda^2}\right] \nonumber \\ &
	+ q \frac{1-\lambda^2}{1+\lambda^2} \gamma_{\vec k},
 \nonumber \\
 c_2(\vec k) =& q \gamma_{\vec k} ,
 \nonumber \\
 c_{3,4} =& \frac{1}{2} \frac{\sqrt{I}}{\sqrt{1+\lambda^2}} \frac{A}{J_\perp} (1\pm\lambda),
 \nonumber \\
 c_5 =& \frac{A}{J_\perp} \frac{\lambda}{1+\lambda^2} .
\end{align}
and
\begin{align}
 c_6(\vec k) = &\frac{1}{1+\lambda^2} \left[ - 4 \frac{A I+\hst}{J_\perp} \lambda     +  (1-\lambda^2)     + 8 \frac{q \lambda^2}{1+\lambda^2}  \right] \nonumber \\ &
   + q \frac{(1-\lambda^2)^2}{(1+\lambda^2)^2}  \gamma_{\vec k}
   \nonumber \\
 c_7(\vec k) = &q  \frac{(1-\lambda^2)^2}{(1+\lambda^2)^2} \gamma_{\vec k}
\end{align}
Note that only the triplon-related terms $c_{1,2,6,7}$ disperse because the hyperfine coupling is purely local. Further, for $\lambda=0$, we see that $c_1=c_6$ and $c_2=c_7$, reflecting the threefold degeneracy of the triplon modes in the SU(2)-symmetric paramagnetic phase realized for $A=0$ and $q<\qcz$.
The Hamiltonian \eqref{eq:h2dim} can be diagonalized with three separate bosonic Bogoliubov transformations for the three components as discussed in the appendix of Ref.~\onlinecite{Wessel05}. The two transverse sectors, $x$ and $y$, yield the same mode energies owing to the residual U(1) symmetry of the ordered state.

\subsection{Order parameter}

An important observable is the electronic staggered magnetization
\begin{align}
	\Mst = \frac{1}{N} \sum_i e^{i \vec Q \vec R_i} (\langle S_{i1,z} \rangle - \langle S_{i2,z} \rangle)
\label{eq:mstdim}
\end{align}
being the order parameter of the electronic quantum phase transition.
Expressed in generalized triplon operators it reads
\begin{align}
	\Mst
	&=  \frac{2 \lambda}{1+\lambda^2} \left(1 - \frac{1}{N}  \sum_{\vec k} \sum_{\alpha \in \{x,y,z\}} \!\! (1+\delta_{\alpha z}) \langle \tilde{t}_{\vec k \alpha}^\dag \tilde{t}_{\vec k  \alpha}^{\phantom \dag} \rangle \right).
\end{align}
The leading term is the product-state contribution $\langle\psi_0|\Mst|\psi_0\rangle$, while the second represents fluctuation corrections from triplon occupation. In our cubic-lattice calculation, we find the fluctuation corrections to be less than 10\% of the leading value. We also note that cubic and higher-order triplon terms lead to additional corrections\cite{Joshi15b} entering the rotation parameter $\lambda$ which have been neglected here.

Without hyperfine coupling, $\Mst$ vanishes for $q\leq \qcz$, while it varies as $\Mst\propto (q-\qcz)^\beta$ for $q\gtrsim \qcz$, with $\beta=1/2$ within our mean-field approximation. At the critical point, a staggered magnetization can be induced by an external staggered field according to $\Mst\propto\hst^{1/\delta}$ with $\delta=3$ at the mean-field level.

\begin{figure}[tb]
	\centering
	\includegraphics[width=1\columnwidth]{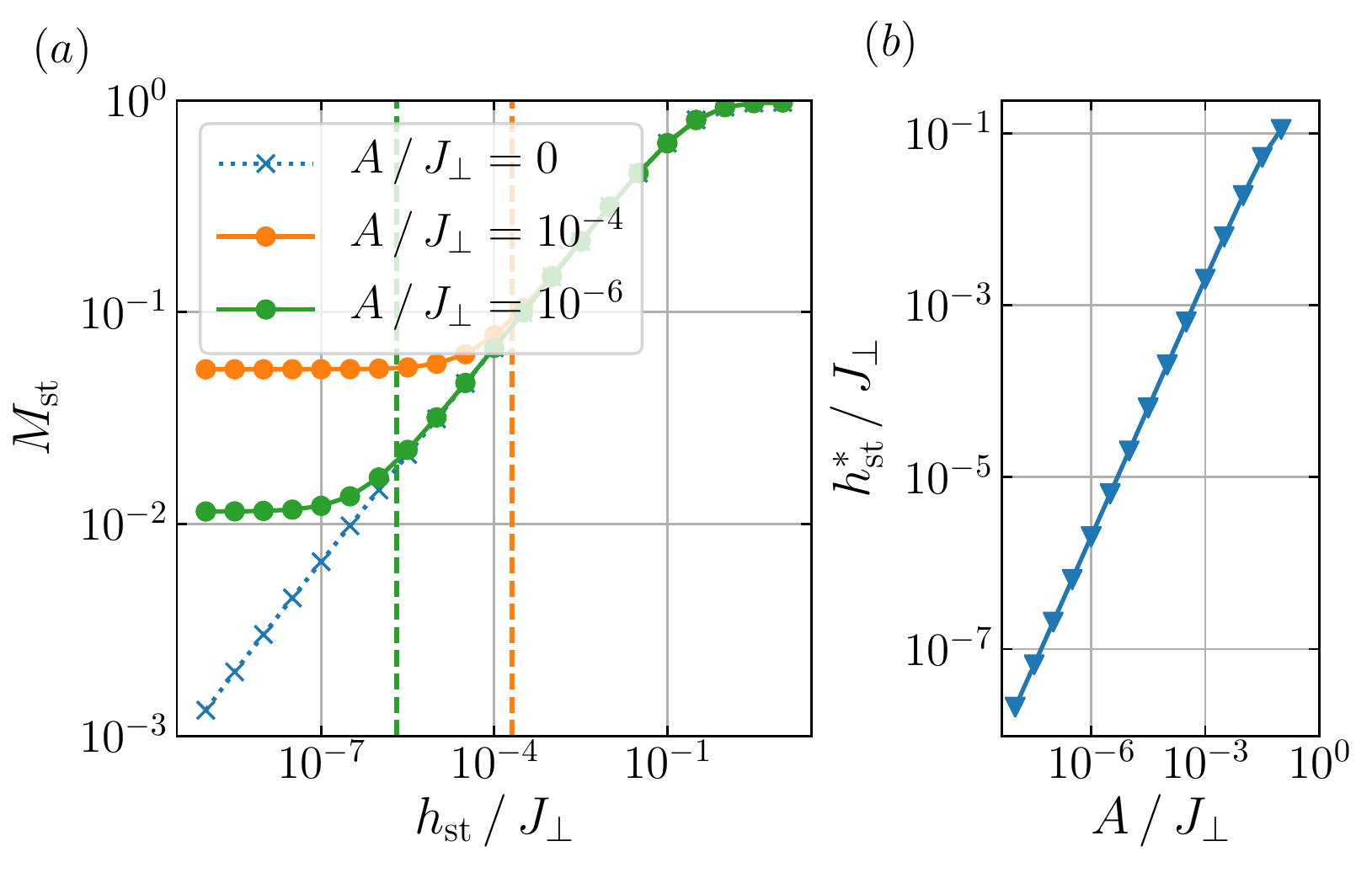}
	\caption{
(a) Electronic staggered magnetization $\Mst$ \eqref{eq:mstdim} of model \eqref{eq:dimers-Ham} as function of the external staggered field $\hst$ for hyperfine coupling $A=0$ and $A \neq 0$.
(b) Crossover field strength $h_{\rm st}^\ast$, where the relative difference of $\Mst$ for $A=0$ and $A\neq0$ is larger than $10\%$, as function of $A$; the position of $h_{\rm st}^\ast$ is also marked by vertical dashed lines in (a).
}
	\label{fig:dimers-MstOfhz}
\end{figure}

In the presence of a finite hyperfine coupling $A$, the staggered magnetization $\Mst$ is generically non-zero, i.e., the QPT is smeared and the critical power laws are cut off. For small $A$ this can be deduced by perturbative arguments.
For $q\leq \qcz$, the hyperfine couplings simply acts as a staggered field, resulting in $\Mst\propto A$ for $q<\qcz$ and $\Mst\propto A^{1/\delta}$ for $q=\qcz$, i.e., at the former QPT. Consequently, the critical power law $\Mst\propto\hst^{1/\delta}$ is cut off at a field $h_{\rm st}^\ast \sim A$, below which $\Mst$ becomes field-independent, as illustrated in Fig.~\ref{fig:dimers-MstOfhz}.
Similarly, the critical power law $\Mst\propto (q-\qcz)^\beta$ is cut-off on a scale $(q-\qcz)^\ast \sim A^{1/(\beta\delta)}$, here $1/(\beta\delta) = 2/3$, as $\Mst$ takes its hyperfine-induced value for $(q-\qcz)$ smaller than $(q-\qcz)^\ast$.

\subsection{Excitation spectrum}
\label{sec:dimers-qps}

\begin{figure}[tb]
	\centering
	\includegraphics[width=\columnwidth]{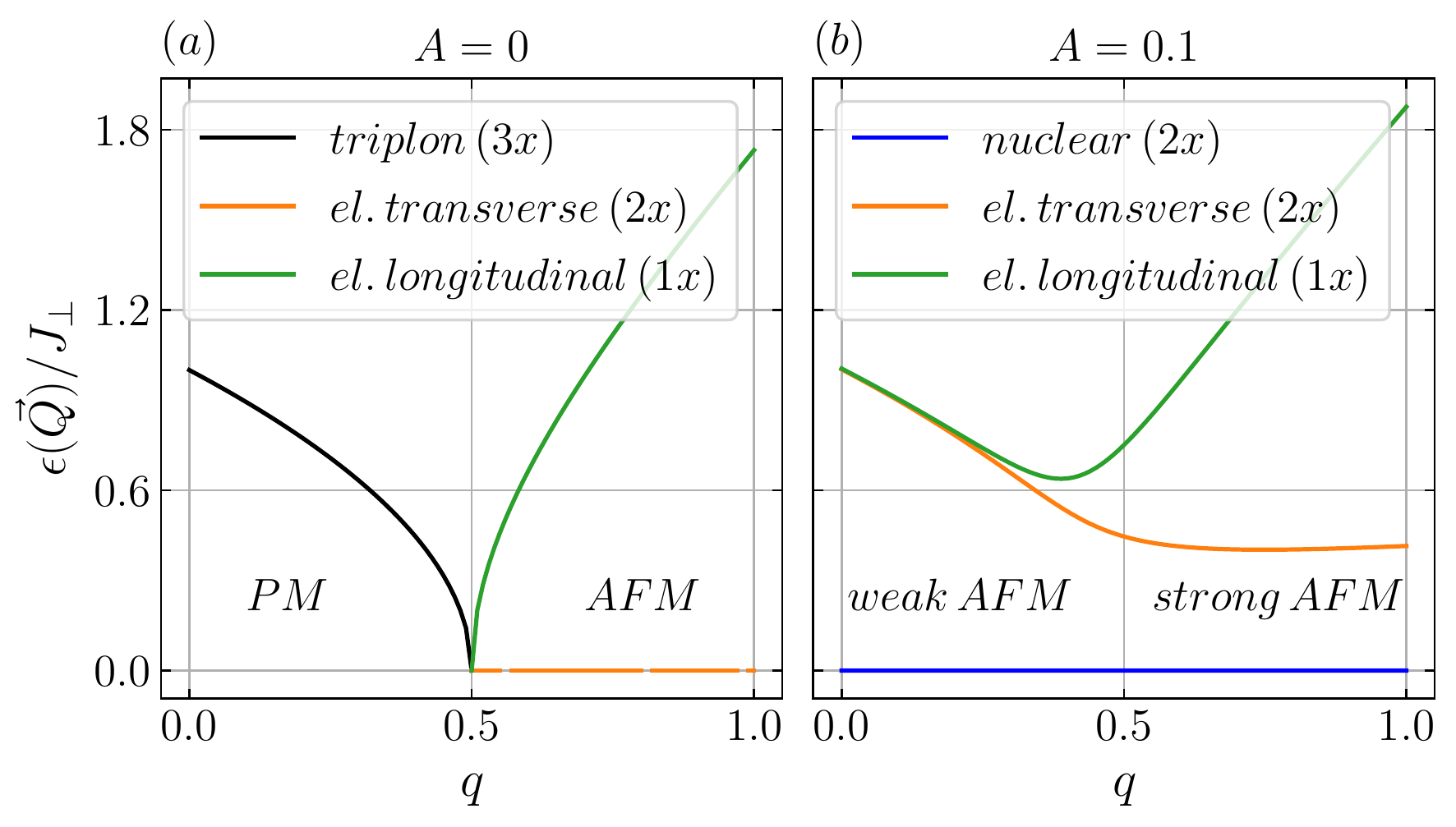}
	\caption{
Mode energies of the dimer model \eqref{eq:dimers-Ham} at the ordering wave vector $\vec k =\vec Q$ (a) without nuclear spins and (b) with finite hyperfine coupling to nuclear spins.
}
	\label{fig:dimers-qpspectrum-overview}
\end{figure}

The excitation spectrum of the coupled system of electronic and nuclear spins determines important crossover scales in the phase diagram shown in Fig.~\ref{fig:dimers-phasediagram}. We recall that the electronic subsystem, in the absence of hyperfine coupling, displays three degenerate and gapped triplon modes in the dimer phase, $q<\qcz$, and two transverse gapless Goldstone modes and a gapped longitudinal (Higgs) mode in the AF phase, $q>\qcz$. The triplon and Higgs gaps vanish upon approaching the QPT at $q=\qcz$ according to $\Delta \propto |q-\qcz|^{\nu z}$ where $\nu$ and $z$ are the correlation-length and dynamic exponents, respectively, with $z=1$ and $\nu=1/2$ at the mean-field level. This is shown in Fig. \ref{fig:dimers-qpspectrum-overview}(a).

Including nuclear spins with a small hyperfine coupling $A$ yields two additional nuclear modes. In the weak antiferromagnet realized for $q<\qcz$, these are the Goldstone modes of the emergent ordered state. Via second-order perturbation theory, their bandwidth is given by $\Wnu \sim A^2/\Delta_0$ with $\Delta_0$ being the triplon gap of the $A=0$ paramagnet. The triplon modes remain gapped, but their degeneracy is lifted due to the coupling to the nuclear modes, reflecting the broken SU(2) symmetry, see Fig. \ref{fig:dimers-qpspectrum-overview}(b).

\begin{figure}[tb]
	\centering
	\includegraphics[width=\columnwidth]{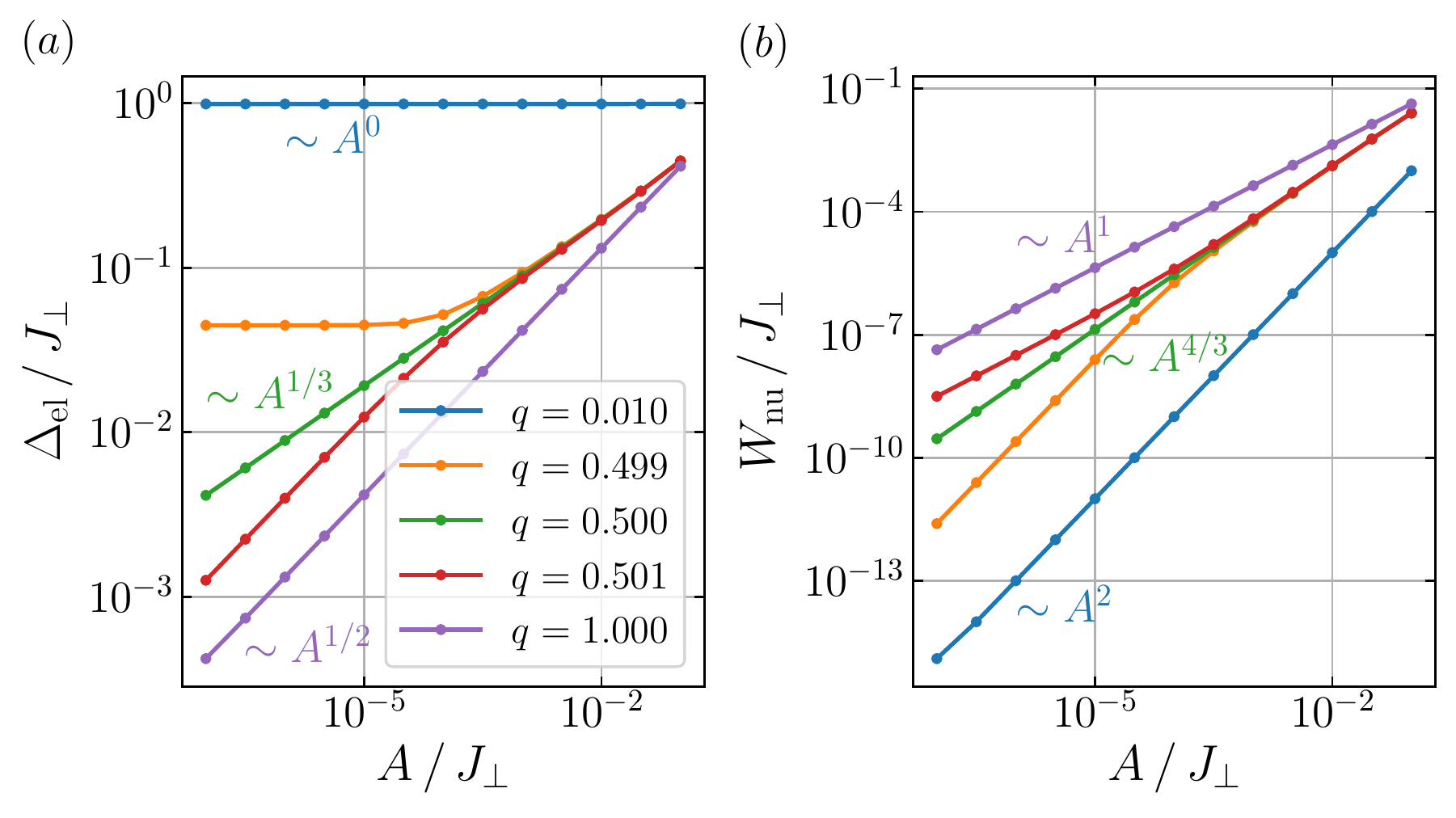}
	\caption{
(a) Gap of the triplon-dominated mode $\Del$ and (b) bandwidth $\Wnu$ of the nuclear-dominated mode, shown as function of hyperfine coupling $A/J_\perp$, for different values of $q$ in model \eqref{eq:dimers-Ham}. Different behavior is seen for $q<\qcz=1/2$, $q=\qcz$, and $q>\qcz$, for details see main text.
}
	\label{fig:dimers-qpspectrum}
\end{figure}

For $q>\qcz$ where AF order is realized also without nuclear spins, the nuclear and electronic modes hybridize strongly for momenta near $\vec k=\vec Q$ where both types of modes are soft for $A=0$. As a result of this hybridization, the lower (nuclear-dominated) modes remain gapless Goldstone modes while the triplon-dominated modes open a gap, the latter scaling as $\Del \sim A^{1/2}$, with the power $1/2$ arising from the structure of the Bogoliubov transformation. Formally, the gap arises from the correction term $\sim A^1$ in the coefficient $c_1$, which gives as a lowest order correction to the electronic energy $\Del \sim \sqrt{c_1^2-c_2^2} \sim  \sqrt{A}$.
The bandwidth of the nuclear mode $\Wnu$ scales as $A$ since it is set by the energy cost of a single nuclear spin flip, given by $A \Mst$. As a result, a distinct gap of excitations arises in the energy range $A<\w<\sqrt{J_\perp A}$, further discussed below.

Finally, at the electronic critical point $q=\qcz$ the triplon-dominated modes (which are critical at $A=0$) develop a gap according to $\Del \sim A^{1/\delta}$, defining a crossover temperature $k_B\Tel$. The $A$ dependence of the gap can be understood as above by noting that, while in the AF phase the correction term to $c_1$ is linear in $A$, at the QPT it is $\sim \lambda^2 \sim A^{2/\delta}$. The nuclear band width grows as $\Wnu \sim A^{(\delta+1)/\delta}$, again for the reason that a single spin flip costs an energy $A \Mst$.

The above behavior is well borne out by our explicit calculations, as illustrated in Fig.~\ref{fig:dimers-qpspectrum}, which shows the behavior of the electronic-mode gap and the nuclear-mode bandwidth as function of $A$. These results also nicely illustrate the crossover occurring near $\qcz$: For $A<|q-\qcz|^{\delta\nu z}$ the above perturbative power laws derived for $q<\qcz$ and $q>\qcz$ hold, whereas $A>|q-\qcz|^{\delta\nu z}$ corresponds to ``quantum critical'' behavior.

\begin{figure}[tb]
	\centering
	\includegraphics[width=0.9\columnwidth]{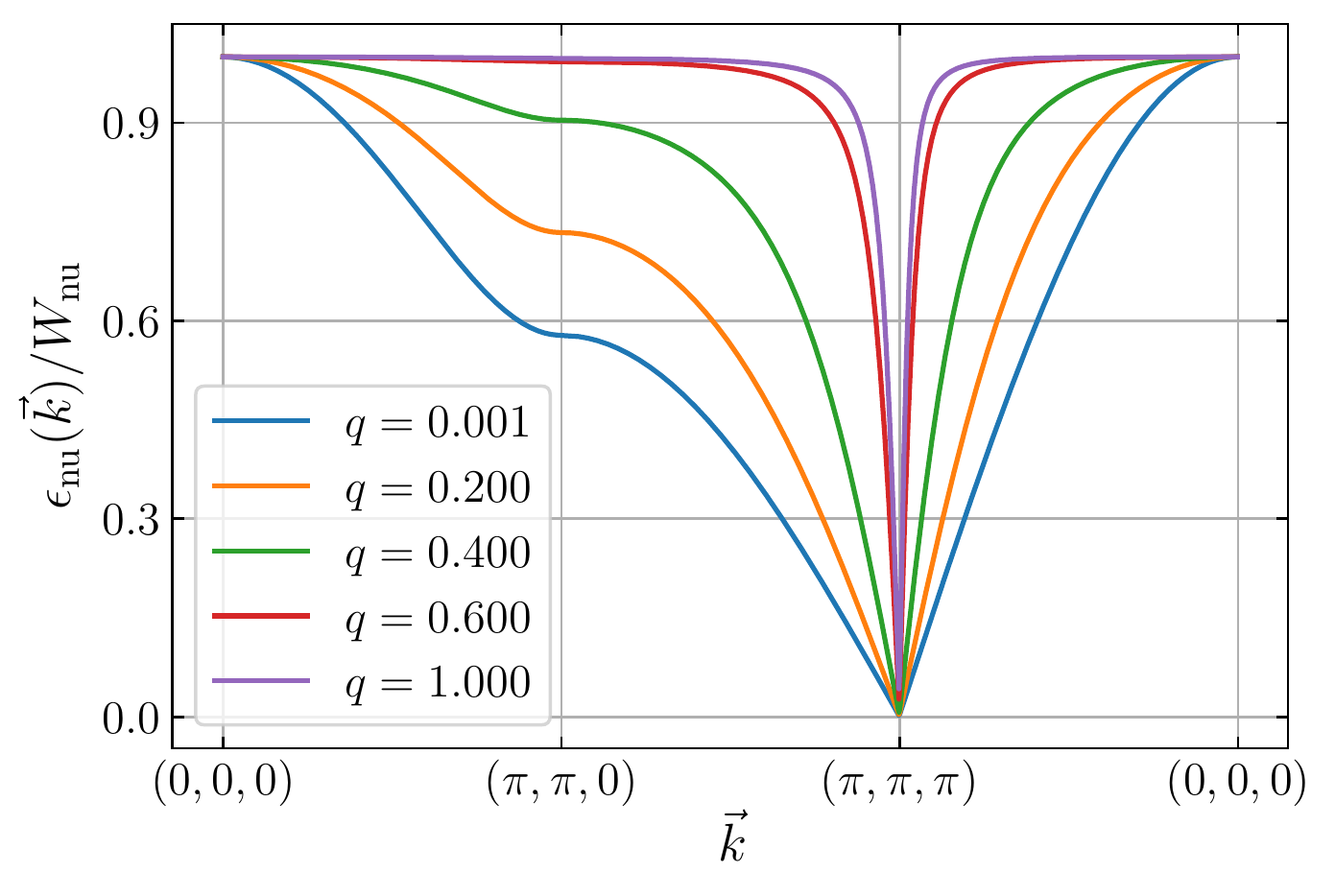}
	\caption{
Dispersion of the nuclear-dominated mode, normalized to its bandwidth, for different $q$ in the coupled-dimer model \eqref{eq:dimers-Ham} at fixed hyperfine coupling $A/J_\perp=10^{-2}$. As $q$ increases, the long-ranged effective interaction between the nuclear spins leads to an increasingly anomalous dispersion.
}
	\label{fig:dimers-nuclMode}
\end{figure}

Our calculation also shows that the dispersion shape of the nuclear-dominated mode changes across the phase diagram, see Fig.~\ref{fig:dimers-nuclMode}. As in Sec.~\ref{sec:ising} the reason is that the interaction between the nuclear spins is mediated by the electronic subsystem. Therefore it is dominated by nearest-neighbor terms if the electronic bandwidth is small compared to the electronic gap, i.e., for $q\ll \qcz$, whereas the interaction becomes increasingly long-ranged with increasing $q$. This gives rise an anomalous dispersion shape for $q\gtrsim \qcz$ where again the Goldstone-mode velocity (in units of the lattice constant) is much larger than the bandwidth, see Fig.~\ref{fig:dimers-nuclMode}.

\begin{figure}
	\centering
	\includegraphics[width=\columnwidth]{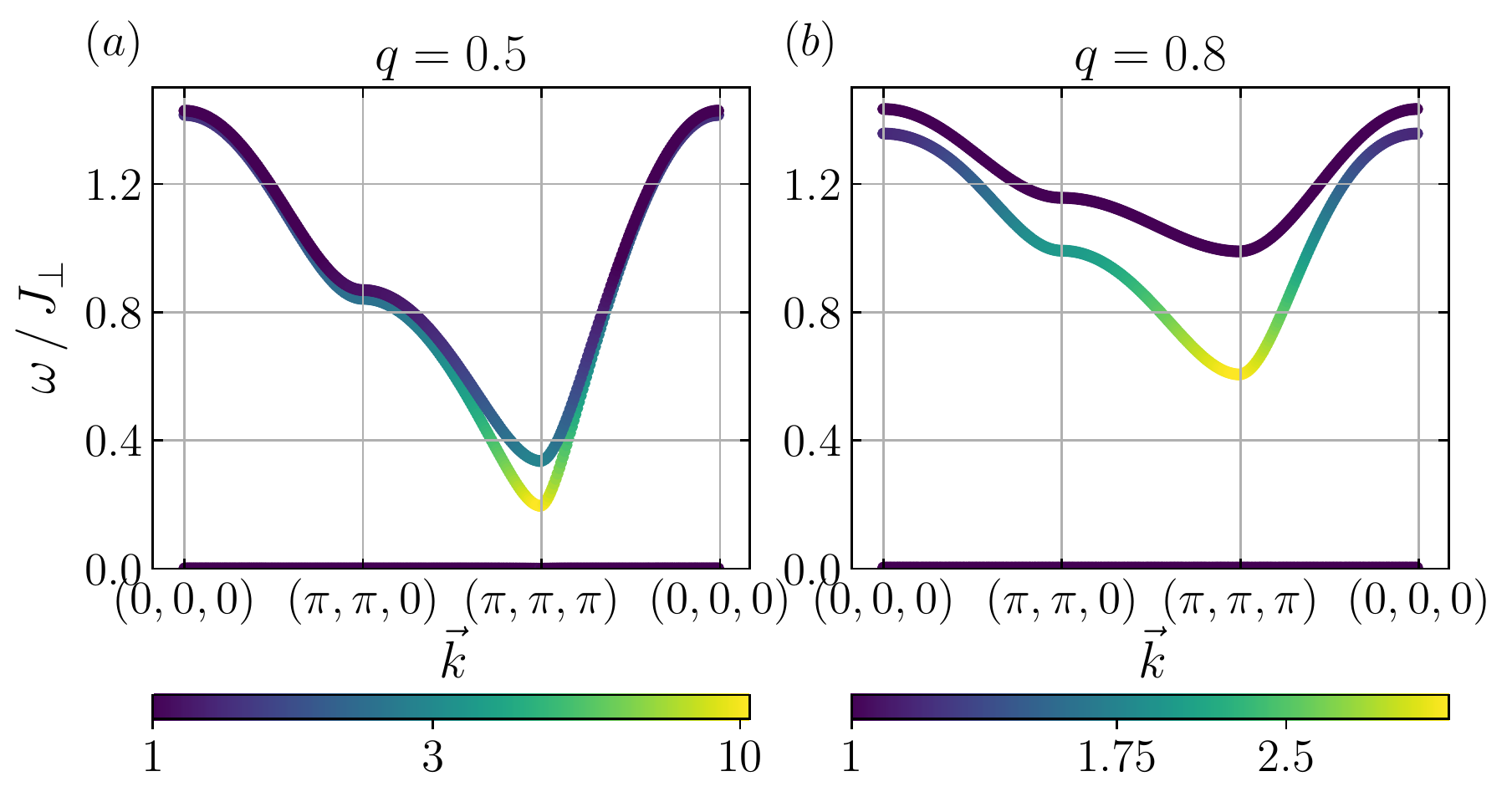}
	\caption{
Imaginary part of the electronic dynamical susceptibility, ${\rm Tr}\,\chi_{\alpha\beta}''(\vec k,\w)$, of the coupled-dimer model \eqref{eq:dimers-Ham} as function of $\vec k$ and $\w$ in the odd channel at $A\,/\,J_\perp=0.01$  and two values of the intradimer coupling $q$. The color code shows the mode weight on a logarithmic scale; the linewidth is artificial.
The hyperfine coupling leads to a finite electronic gap both
(a) at the former critical point $\qcz=0.5$ and
(b) in the former AF phase $q>\qcz$.
The degeneracy of the electronic modes, present for $q\leq\qcz$ and $A=0$, is lifted for finite $A$, see also Fig.~\ref{fig:dimers-qpspectrum-overview}. The weight of the nuclear mode is significant only very close to the ordering wave vector, comparable to the inset of Fig. \ref{fig:ising-susc}.
}
	\label{fig:dimers-susc1}
\end{figure}

From the quasiparticle spectrum we calculate the dynamical susceptibility of the electronic spins as in Eq.~\eqref{eq:chi}. For the coupled-dimer model it is useful to distinguish even and odd channels, where $S_{e,\alpha} = S_{i1,\alpha} + S_{i2,\alpha}$ and $S_{o,\alpha} = S_{i1,\alpha} - S_{i2,\alpha}$, respectively. As before, we restrict ourselves to the single-mode contributions to $\chi''(\vec k,\w)$.
Sample results for the trace of the susceptibility tensor, as measurable in unpolarized neutron scattering, are shown in Fig.~\ref{fig:dimers-susc1}. For realistic values of the hyperfine coupling the nuclear-dominated mode lies at very low energies, which likely makes it difficult to resolve in neutron-scattering experiments. It carries considerable weight only close to the ordering vector $\vec Q$, while at other wavevectors the hybridization between the electronic and nuclear modes is very weak. The triplon gap is significant even at small $A$ -- recall that this scales as $A^{1/2}$ for $q>\qcz$ and $A^{1/\delta}$ for $q=\qcz$ -- such that incomplete electronic mode softening is an accessible indicator of hyperfine effects, as already noted above.
However, distinguishing smeared and shifted phase transition invariably requires measurements at nuclear-spin energy scales and temperatures.

\subsection{Thermodynamics}

Our results allow to deduce the finite-temperature phase diagram of the Heisenberg model \eqref{eq:dimers-Ham} in the presence of small hyperfine coupling $A$, shown in Fig.~\ref{fig:dimers-phasediagram}.
In $d>2$ space dimensions the antiferromagnetic order persists at finite temperature, with the N\'eel temperature $\TN$ scaling with the electronic coupling $J$ for $q\gg \qcz$. In contrast, for $q\leq \qcz$ the order is hyperfine-induced, and $\TN$ can be estimated from the bandwidth of the nuclear-dominated excitation modes. Hence, $\TN\propto A^2/\Delta_0$ for $q\ll \qcz$, while $\TN\propto A^{(\delta+1)/\delta}$ at $q=\qcz$. For small $A$, this implies a strong enhancement of the hyperfine-induced $\TN$ near the electronic quantum critical point compared to the paramagnetic phase.

Near $\qcz$, electronic quantum criticality is visible at elevated temperatures only, while it is cut off for $T$ smaller than $\Tel$ corresponding to the hyperfine-induced gap which scales as $A^{1/\delta}$. This implies, for instance, that the specific heat $C(T)$ will follow its critical power law $C(T)\propto T^{d/z}$ (valid at and below $d_c^+$) for $T>\Tel$, while $C(T)$ drops when cooling below $\Tel$, with nuclear contributions only appearing near (and below) the N\'eel temperature $\TN\propto A^{(\delta+1)/\delta}$.


\begin{figure*}[!tb]
	\centering
	\includegraphics[width=\textwidth]{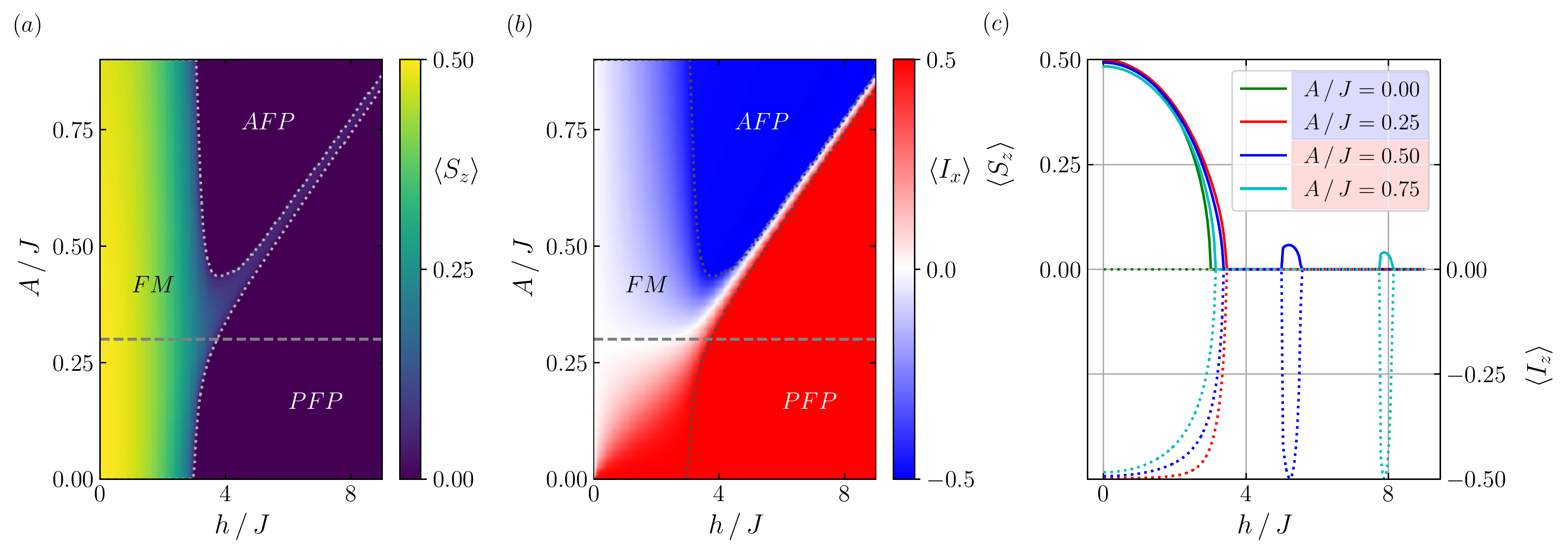}
	\caption{
Results for the transverse-field Ising model illustrating the distinct behavior in regimes II and III, here for $\tilde g_N=0.05$.
(a,b) Phase diagram as function of transverse field $h$ and hyperfine coupling $A$, with the color codes corresponding to (a) the $z$-component $\langle S_z\rangle$ of the electronic spin and (b) the $x$-component $\langle I_x\rangle$ of the nuclear spin. Dotted lines are phase boundaries; the horizontal dashed line corresponds to $A=2d\tilde{g}_N J$ separating regimes II and III.
At low fields, the electronic spins exhibit ferromagnetic order (FM). At high fields, two field-polarized phases with nuclear spins parallel (PFP) or antiparallel (AFP) to the electronic spins occur. AFP and PFP are separated by a corridor of FM order mainly carried by nuclear spins at $A S \approx \tilde g_N h_x$.
(c) $\langle S_z\rangle$ (full) and $\langle I_z\rangle$ (dotted) as function of field for different fixed $A/J$.
}
	\label{fig:ising-recurrorder}
\end{figure*}

\section{Hyperfine-induced additional phases}
\label{sec:novel}

So far the discussion focused on the regime of small hyperfine coupling and on unfrustrated systems. If the hyperfine coupling becomes comparable to other energy scales in the system, then new phenomena can occur -- this is the subject of this section. Specifics of frustrated systems will be discussed in Sec.~\ref{sec:exotic} below.

Given that hyperfine couplings are generically smaller than electronic couplings, the most obvious small energy scale the hyperfine coupling can compete with is the nuclear Zeeman energy. To illustrate this, we will return to the transverse-field Ising model in Eq.~\eqref{eq:ising-Ham} of Sec.~\ref{sec:ising}. Here, we will focus mostly on regime III where $A \gtrsim 2d \tilde g_N J $, i.e., the hyperfine coupling is comparable to or larger than the nuclear Zeeman energy near the transition.

At small applied field a large (antiferromagnetic) hyperfine coupling enforces antiparallel electronic and nuclear spins; in regime III this continues to be true across the transition. In contrast, in the asymptotic high-field limit, both spin species are forced to align with the external field, such that they are parallel. This requires at least one additional quantum phase transition with increasing applied field.

For a detailed modelling, we solve Eq.~\eqref{eq:ising-Ham} in a modified mean-field approximation, where the hyperfine coupling is treated exactly for $S=I=1/2$, i.e., we resort to a product reference state of the form $|\psi_0\rangle = \prod_i |\tilde s\rangle_i$, where $|\tilde s\rangle_i$ is an arbitrary state from the four-dimensional Hilbert space formed by $S_i$ and $I_i$ which also includes the possibility of singlet formation between electronic and nuclear spins.
Minimizing $\Eps = \langle \psi_0 | \mathcal{H}^{\rm TI} | \psi_0 \rangle$ is equivalent to solving the mean-field Hamiltonian
\begin{align}
	\HMF^{\rm TI} = &- 2d J  \langle S_z \rangle S_z - \vec h (\vec S + \tilde{g}_N \vec I) + A \vec S \cdot \vec I
\end{align}
at zero temperature.

Sample results are shown Fig.~\ref{fig:ising-recurrorder}, where we have chosen a large $\tilde g_N=0.05$ for illustration purposes. We find that, remarkably, the field-driven change from antiparallel (AFP) to parallel (PFP) field-polarized electronic and nuclear spins proceeds via two QPTs (instead of a single one), with an intermediate ordered phase occurring (!). This can be rationalized as follows: For $A S \approx \tilde g_N h$ the mean field acting on the nuclear spins becomes small, such that their behavior is determined by their indirect interaction via the electronic subsystem. As a result, weak ferromagnetic order carried primarily by the nuclear spins occurs in a narrow field window whose width decreases with increasing $A$.

Interestingly, such behavior is expected in the paradigmatic transverse-field Ising material {\lhf}, Sec.~\ref{sec:LHF}, which turns out to be located in regime III. Quantitative modelling\cite{ourLHFpaper} shows that the nuclear spins are antiparallel to the field near the quantum phase transition at $5.1$\,T, and the double transition to field-aligned nuclear spins is expected at a field of roughly $300\,$T.


\section{Frustration and exotic transitions}
\label{sec:exotic}

In this section, we provide a perspective on the effects of nuclear spins on quantum criticality beyond conventional order-disorder transitions. As before, we restrict ourselves to local-moment Mott insulators where frustration can induce both non-trivial phases such as fractionalized spin liquids and transitions beyond the Landau-Ginzburg-Wilson paradigm.

\subsection{Spin liquids and hyperfine coupling}

We start our qualitative considerations by discussing the fate of spin-liquid states upon inclusion of nuclear spin degrees of freedom. Given that time-reversal symmetry is unbroken in a spin liquid, the mechanism of nuclear spin order discussed in Sec.~\ref{sec:dimers} can modify or destabilize a given spin liquid.

For the honeycomb-lattice Kitaev model, \cite{Kitaev06} the hyperfine-induced effective interaction between the nuclear spins is also of Kitaev type, with only nearest-neighbor terms, and therefore the $Z_2$ spin liquid can be expected to be stable in the presence of nuclear spins.
This is possibly different for the nearest-neighbor Heisenberg model on the kagome lattice \cite{Depenbrock12, He17} which is a candidate for a $Z_2$ spin liquid -- either gapped or gapless -- as well. The effect of nuclear spins in this model has been discussed before\cite{Legg19}, with the conclusion that nuclear spin order is not stable. However, this analysis was based on a low-order estimate of the electronic spin susceptibility which was considered to be momentum-independent. It is likely that the full electronic susceptibility displays momentum dependence, arising, e.g., from the presence of Dirac cones in the spectrum, such that a stable nuclear spin order is induced at low temperatures which in turn induces weak order of the electron spins.
Similar considerations apply to the quantum spin liquid realized in the $J_1$-$J_2$ triangular lattice.\cite{Hu15, Liu20}
Consequently, electronic quantum criticality arising from transitions out of such spin liquids is likely to be cut off by the ordering scale arising from nuclear spins, not unlike the situation in Sec.~\ref{sec:dimers}.

For classical spin liquids, such as spin ice, the Heisenberg coupling to nuclear spins induces quantum dynamics which eventually quenches the finite entropy accompanying the degenerate classical manifold. In a realistic setting such as Ho$_2$Ti$_2$O$_7$ the hyperfine-induced quantum dynamics competes or cooperates with the (weak) intrinsic quantum dynamics of the electronic spin system.
Parenthetically, we note that highly interesting behavior arising from nuclear spins has also been detected in the kagome ice material Ho$_3$Mg$_2$Sb$_3$O$_{14}$.\cite{Dun20} Here, the peculiar form of the crystalline electric field leads to a weakly split non-Kramers doublet, and this intrinsic splitting competes with dipolar and exchange couplings as well as with hyperfine interactions.

\subsection{Deconfined criticality of VBS-N\'{e}el type and hyperfine coupling}

As a representative for exotic transitions, we consider deconfined quantum criticality. Deconfined criticality refers to Landau-forbidden continuous quantum transitions between two symmetry breaking states where the critical point features fractionalized excitations.\cite{Senthil04, Senthil04PRB, Vishwanath04} A paradigmatic example is the transition between a N\'eel antiferromagnet and a valence-bond solid as realized, e.g., in the so-called $J$-$Q$ model on the square lattice. \cite{Sandvik07}

Upon adding hyperfine-coupled nuclear spins, the valence-bond solid will become a weak antiferromagnet, as the nuclear spins experience an interaction favoring N\'eel-type order. This weak spin order coexists with spontaneous dimerization. As a result, the former deconfined QPT will be replaced by a QPT involving the onset of dimerization only, as both phases display the same type of magnetic order. Hence, the physics of the deconfined QCP is cut off at low energies and temperatures where it is replaced by that of a more conventional $Z_4$ QCP.


\section{Conclusions and outlook}
\label{sec:concl}

In the context of magnetic quantum phase transitions in solids, we have discussed the fate of electronic quantum criticality under the influence of nuclear spins. Their hyperfine coupling to the electrons can modify or even destroy the criticality of the electronic system.
If time-reversal symmetry is broken in both electronic phases, e.g., by an applied magnetic field, the hyperfine coupling is a regular perturbation and the quantum phase transition is shifted. At low temperature, a regime of nuclear-spin quantum criticality may emerge.
If instead the disordered electronic phase is time-reversal symmetric, it may be unstable to weak magnetic order at low temperatures, carried by nuclear spins, and the quantum phase transition is turned into a crossover.
We have considered explicit examples for both cases and evaluated the relevant crossover scales below which electronic quantum criticality gets modified. We have also discussed situations with hyperfine-induced additional transitions.

Systems with magnetic 4f electrons are particularly suitable to detect the phenomena discussed in this paper, as they feature small exchange interactions in the Kelvin range and potentially significant hyperfine coupling. Both thermodynamic measurements and inelastic neutron scattering are suitable to detect the crossover scales predicted here. This has partially been done for {\lhf}, and our work prompts for more detailed experimental studies.

In terms of an outlook, it is worth mentioning a number of related experimental findings which point a particular role of nuclear spins in regimes with strong quantum fluctuations.
One case in point is CeRu$_2$Si$_2$ where very unusual thermodynamic behavior has been reported\cite{Yoshida08} at temperatures below $50$\,mK which has been attributed to an additional quantum critical point. Given that both $^{99}$Ru and $^{101}$Ru isotopes have a large nuclear spin, nuclear magnetism is a plausible candidate, but a detailed modelling is lacking.
Signatures of the particularly effective coupling of nuclear and electronic spins have also been seen\cite{Kim09} in calorimetric measurements on Cr(diethylenetriamine)(O$_2$)$_2 \cdot$H$_2$O which contains a large number of hydrogen nuclear spins.


\acknowledgments

We thank C. Duvinage, L. Janssen, C. Pfleiderer, and A. Wendl for discussions and collaborations on related work. MV also thanks T. Ulbricht for an early
collaboration on a related topic.
Financial support from the Deutsche Forschungsgemeinschaft through SFB 1143 (Project No. 247310070) and the W\"urzburg-Dresden Cluster of Excellence on Complexity and Topology in Quantum Matter -- \textit{ct.qmat} (EXC 2147, Project No. 390858490) is gratefully acknowledged.


\appendix

\section{Derivation of the excitation spectrum of the Ising model}
\label{app:isingcalc}

The classical reference state is defined by the angles $\theta$ and $\phi$ specifying the directions of the electronic and nuclear spins as shown in Fig.~\eqref{fig:ising-model}. To apply spin-wave theory, we formally rotate the spins to a new frame of reference $\vec S_i = U_{\rm el} \tilde{\vec S}_i$ and $\vec I_i = U_{\rm nu} \tilde{\vec I}_i$ in which they are aligned with the $z$-axis. For the electronic spins the site-independent rotation matrix reads
\begin{align}
	U_{\rm el} =
\begin{pmatrix}
\cos \theta & 0 & \sin \theta  \\
0 & 1 & 0  \\
-\sin \theta & 0 & \cos \theta
\end{pmatrix},
\end{align}
and the rotation for nuclear spins is obtained by replacing $\theta\to-\phi$. For generality we take a possible longitudinal field into account, $\vec h =(h_x,0,h_z)^T$, and consider arbitrary spin sizes $S$ and $I$.

These representations of the spins in a rotated frame of reference can be inserted into the model Hamiltonian 
\begin{widetext}
Eq.~\eqref{eq:ising-Ham} and with the abbreviations $\alpha = \cos\theta \cos\phi - \sin\theta \sin\phi$, $\beta = \cos\theta \sin\phi + \sin\theta \cos\phi$ one obtains
\begin{align}
	\mathcal{H}^{\rm TI} =
	&- J \sum_{\langle i j \rangle} \left[ \sin^2 \theta \, \tilde S_{ix} \tilde S_{jx}  +\cos^2 \theta \, \tilde S_{iz} \tilde S_{jz}
	-\sin \theta \cos \theta (\tilde S_{ix}\tilde S_{jz} + \tilde S_{iz} \tilde S_{jx} )  \right]
	\nonumber \\
	& - \sum_i  \left[ \left(  h_x \cos  \theta  -h_z \sin \theta \right) \tilde S_{ix}
	+	\left(h_x \sin \theta + h_z \cos \theta  \right) \tilde S_{iz} \right]
	\nonumber \\
	& - \tilde g_N \sum_i  \left[ \left(  h_x \cos \phi  +h_z \sin \phi  \right) \tilde I_{ix}
	+ \left(-h_x \sin \phi  + h_z \cos \phi \right) \tilde I_{iz} \right]
	\nonumber \\
	&+ A \sum_i \left[
	 \alpha \tilde S_{ix} \tilde I_{ix}
 	- \beta \tilde S_{ix} \tilde I_{iz}
	+ \beta \tilde S_{iz} \tilde I_{ix}	
	+ \alpha \tilde S_{iz} \tilde I_{iz}
	+  \tilde S_{iy} \tilde I_{iy}  \right]
\end{align}
We now represent the spin operators using the standard Holstein-Primakoff representation \cite{Kittel,Stancil},
$\tilde S_{iz} = S - a_i^\dag a_i$,
$\tilde S_{i+} = (2S - a_i^\dag a_i)^{1/2} a_i$ and
$\tilde I_{iz} = -I + b_i^\dag b_i$,
$\tilde I_{i-} = (2I - b_i^\dag b_i)^{1/2} b_i$.

Insertion gives a constant energy contribution and a linear term, whose coefficient vanishes if the angles $\theta$ and $\phi$ of the reference state are chosen correctly, see main text. We neglect interaction effects, so that only the quadratic term
remains.
To solve it, we first go to momentum space
\begin{align}
 \mu_i = \frac{1}{\sqrt{N}} \sum_{\vec k} e^{i \vec k \vec r_i} \mu_{\vec k}
 \label{eq:fourierdef}
\end{align}
where $\mu$ stands for the two types of Holstein-Primakoff bosons $a,b$. We introduce the sum over nearest neighbors
\begin{align}
  \gamma_{\vec k} = \frac{1}{2d} \sum_{j \in {\rm NN}_i} e^{i \vec k (\vec r_j - \vec r_i)},
  \label{eq:gammadef}
\end{align}
which e.g. for the cubic lattice is $\gamma_{\vec k} = (\cos k_x + \cos k_y + \cos k_z)/3$.
Then the Hamiltonian can be written as
\begin{align}
\mathcal{H}_2^{\rm TI}
=& - J d \frac{S}{2} \sin^2\theta \, \sum_{\vec k} \left[ \gamma_{\vec k} (a_{-\vec k}^{\phantom \dag }a_{\vec k}^{\phantom \dag } + a_{\vec k}^\dag a_{\vec k}^{\phantom \dag } )
%
	+ \gamma_{\vec k}^\ast( a_{\vec k}^{\phantom \dag} a_{\vec k}^\dag + a_{-\vec k}^\dag a_{\vec k}^\dag ) \right]
	\nonumber \\
&+\left[ 2Jd S \cos^2 \theta  + h_x \sin \theta + h_z \cos \theta +  A \alpha I \right] \sum_{\vec k} a_{\vec k}^\dag a_{\vec k}
 +\left[\tilde g_N h_x \sin \phi - \tilde g_N h_z \cos \phi +  A \alpha S \right] \sum_{\vec k} b_{\vec k}^\dag b_{\vec k}
	\nonumber \\ &
+ A \frac{\sqrt{IS}}{2} \sum_{\vec k} \left[(\alpha-1) a_{\vec k}^{\phantom \dag } b_{\vec k}^\dag + (\alpha+1)a_{-\vec k}^{\phantom \dag } b_{\vec k}^{\phantom \dag }
+ (\alpha+1)a_{-\vec k}^\dag b_{\vec k}^\dag + (\alpha-1)a_{\vec k}^\dag b_{\vec k}^{\phantom \dag } \right]
\end{align}

This quadratic Hamiltonian can be diagonalized with a Bogoliubov transformation. The results for $h_x=h$ and $h_z=0$ are discussed in the main text.


\section{Derivation of the excitation spectrum of the coupled-dimer model}
\label{app:dimerscalc}

To derive the Hamiltonian describing coupled electronic and nuclear excitations, we start with triplon excitations. We follow the procedure of Sec. II A in Ref.~\onlinecite{Joshi15b} to rewrite the spin operators of the full Hamiltonian in Eq.~\eqref{eq:dimers-Ham} in terms of triplon operators.
The triplon expressions corresponding to the electronic coupling terms $J_\perp$ and $J_\parallel$ are given by the corresponding terms in Eqs.~(24-26) in Ref.~\onlinecite{Joshi15b}, but to describe the hyperfine coupling we need to include a general magnetic field instead of only the staggered field $\hst$ considered in the reference. This general Zeeman term can be rewritten with triplon operators as
\begin{align}
   - \sum_i (\vec h_{i1} &\cdot \vec S_{i1} + \vec h_{i2} \cdot  \vec S_{i2})
  \nonumber \\
  =  &-\frac{\lambda}{1+\lambda^2} \sum_i e^{i \vec Q \vec r_i} (h_{i1,z} - h_{i2,z})
     - \frac{1}{2} \sum_i  \left\{ \left( \frac{h_{i1,x}-h_{i2,x}}{\sqrt{1+\lambda^2}} + \frac{- i \lambda_i (h_{i1,y}+h_{i2,y})}{\sqrt{1+\lambda^2}} \right) \tilde t_{ix}
    \right. \nonumber \\ & \left.
   + \left( \frac{+ i \lambda_i (h_{i1,x}+h_{i2,x})}{\sqrt{1+\lambda^2}} + \frac{h_{i1,y}-h_{i2,y}}{\sqrt{1+\lambda^2}} \right) \tilde t_{iy}
      + \frac{1-\lambda^2}{1+\lambda^2} (h_{i1,z}-h_{i2,z}) \tilde t_{iz}
      + h.c. \right\}
   \nonumber \\
    & + \frac{\lambda}{1+\lambda^2} \sum_i  \sum_{\beta=1}^3 e^{i \vec Q \vec r_i} (h_{i1,z}-h_{i2,z}) (1+\delta_{\beta z}) \tilde t_{i\beta}^\dag  \tilde t_{i\beta}
     - \frac{1}{2} \sum_i  \left\{ - \left( h_{i1,z}+h_{i2,z} \right) \tilde t_{ix}^\dag \tilde t_{iy}
    \right. \nonumber \\ & \left.
    + \left( \frac{\lambda_i (-h_{i1,x}+h_{i2,x})}{\sqrt{1+\lambda^2}} + \frac{-i (h_{i1,y}+h_{i2,y})}{\sqrt{1+\lambda^2}} \right) \tilde t_{ix}^\dag \tilde t_{iz}
    + \left( \frac{i (h_{i1,x}+h_{i2,x})}{\sqrt{1+\lambda^2}} + \frac{\lambda_i (-h_{i1,y}+h_{i2,y})}{\sqrt{1+\lambda^2}} \right) \tilde t_{iy}^\dag \tilde t_{iz}
    + h.c. \right\}
   \nonumber \\
   & + \text{terms with $\geq 3$ triplon operators}
   \label{eq:fieldTriplon}
\end{align}
We are only interested in terms with $2$ or less triplon operators because we will later employ a non-interacting (harmonic) approximation.

To describe the coupling to the nuclear spins in lowest-order spin-wave theory, we first go to a local reference frame $(I_x, I_y, I_z)\rightarrow (\tilde I_x, e^{i \vec Q r_i} \tilde I_y, e^{i \vec Q r_i} \tilde I_z)$. In the reference state, the rotated spins $\tilde{\vec I}_{im}$ within each layer all point in the same direction, $\tilde{\vec I}_{i1} = (0,0,-I)$ and $\tilde{\vec I}_{i2} = (0,0,I)$ for all sites $i$. In this way, we can use a unit cell of one dimer and two nuclear spins instead of the doubled unit cell of the ordered state, which simplifies the calculation. 

Fluctuations around this state can be parameterized by Holstein-Primakoff bosons \cite{Kittel, Stancil} for each constituent $1$ and $2$ of the dimer as
$\tilde I_{i1,z} = -I + a_i^\dag a_i$,
$\tilde I_{i1,-} = (2I - a_i^\dag a_i)^{1/2} a_i$ and
$\tilde I_{i2,z} = I - b_i^\dag b_i$,
$\tilde I_{i2,+} = (2I - b_i^\dag b_i)^{1/2} b_i$.
We only keep the lowest non-constant order.
Thus we need to formally replace the field $\vec h_{im} = - A \vec I_{im} - (-1)^{m} e^{i \vec Q \vec r_i} \hst \vec e_z$ which couples to the triplons by
\begin{align}
 \vec h_{i1} &= \left( A\frac{\sqrt{2I}}{2} (a_i^\dag+a_i) , -iA e^{i \vec Q r_i} \frac{\sqrt{2I}}{2} (a_i^\dag-a_i),
  e^{i \vec Q r_i}(-AI + Aa_i^\dag a_i - \hst) \right)
 \nonumber \\
 \vec h_{i2} &= \left( A \frac{\sqrt{2I}}{2} (b_i+b_i^\dag) , -i A e^{i \vec Q r_i}\frac{\sqrt{2I}}{2} (b_i-b_i^\dag),
 e^{i \vec Q r_i}(AI - Ab_i^\dag b_i + \hst )\right)
 \label{eq:fieldHP}
\end{align}
Note the sign factors $e^{i \vec Q r_i}$ in the $y$ and $z$ components which come from reversing the transformation $\vec I \rightarrow \tilde{\vec I}$.

Inserting Eq.~\eqref{eq:fieldHP} into Eq.~\eqref{eq:fieldTriplon}, we can include the purely electronic terms as in Ref.~\onlinecite{Joshi15b}. In the resulting Hamiltonian, the constant term represents the product-state energy, and terms linear in boson operators are absent because the product state corresponds an energy minimum for a properly chosen rotation parameter $\lambda(q,A)$. We then focus on bilinear boson terms, while higher-order contributions reflecting boson interactions, are neglected.
We note that only the combinations $c_i = (a_i - b_i)/\sqrt{2}$ and $d_i = e^{i\vec Q \vec r_i}(a_i + b_i)/\sqrt{2}$ of the nuclear bosons appear, so we change to this basis of the nuclear Hilbert bosons.

The Hamiltonian can be simplified by applying a Fourier transform as defined in Eq.~\eqref{eq:fourierdef} where $\mu$ now stands for the five boson species $c, d, \tilde t_x, \tilde t_y,\tilde t_z$. This yields
\begin{align}
 \mathcal{H}_2^{\rm CD} =
     &+ A \frac{\lambda}{1+\lambda^2} \sum_{\vec k} ( c_{\vec k }^\dag c_{\vec k } + d_{\vec k }^\dag d_{\vec k })
     \nonumber \\
 &+ \frac{1}{4} \frac{\sqrt{2I} A }{\sqrt{1+\lambda^2}} \sum_{\vec k}  \left\{ \left[
    (1+\lambda) c_{- \vec k}+ (1-\lambda) c_{\vec k}^\dag \right]  \tilde t_{\vec k x}
         - i \left[ (1+\lambda)  d_{-\vec k} - (1-\lambda) d_{\vec k }^\dag  \right] \tilde t_{\vec k y}
      + h.c. \right\}
  \nonumber \\
 &+ \frac{1}{1+\lambda^2}  \sum_{\beta=1}^3 \left[ - 2 \lambda \left(A I+\hst\right) (1+\delta_{\beta z})
    + J_\perp  \left( 1-\lambda^2 \delta_{\beta z} \right)
    + 4 J_\parallel z \frac{\lambda^2}{1+\lambda^2} (1+\delta_{\beta z}) \right]
    \sum_{\vec k} \tilde t_{\vec k \beta}^\dag  \tilde t_{\vec k \beta}
 \nonumber \\
   & + J_\parallel z \frac{1-\lambda^2}{1+\lambda^2} \sum_{\beta=1}^3 \left[ 1 -\frac{2\lambda^2}{1+\lambda^2}\delta_{\beta z}\right]
    \sum_{\vec k} \gamma_{\vec k} \; \tilde t_{\vec k \beta}^\dag  \tilde t_{\vec k \beta}
 + \frac{J_\parallel z}{2} \sum_{\beta=1}^3 \left[ 1 -\frac{4\lambda^2}{(1+\lambda^2)^2}\delta_{\beta z}\right]  \sum_{\vec k} \left( \gamma^\ast_{\vec k} \; \tilde t_{\vec k \beta}^\dag  \tilde t_{-\vec k \beta}^\dag  + h.c. \right)
\end{align}
with the sum over nearest neighbors $\gamma_{\vec k}$ defined in Eq.~\eqref{eq:gammadef}.

\end{widetext}

The Hamiltonian is quadratic, so it can be diagonalized with a bosonic Bogoliubov transformation. It couples only $\tilde t_{\vec k x}$ with $c_{\vec k}$ and $\tilde t_{\vec k y}$ with $d_{\vec k}$, while $\tilde t_{\vec k z}$ remains separate, so the block-diagonal form of the Hamiltonian can be used to separate the system of $5$ coupled boson species into a simpler system of $(2+2+1)$ coupled bosons. The results of the diagonalization are discussed in the main text.


\bibliographystyle{apsrev4-1}
\bibliography{references_qptnucl}

\end{document}